\documentclass[pra,showpacs,twocolumn]{revtex4}
  \usepackage{amsmath}
  \usepackage{amssymb}
  \usepackage{bm}
  \usepackage{hyperref}
\usepackage{natbib}

  \usepackage{graphicx}
  \def\dfrac{\displaystyle\frac}

 \newcommand {\e}{{\rm e}}
\newcommand {\rmi}{{\rm i}}
\newcommand {\rmd}{{\rm d}}

\renewcommand {\Im}{\mathop{\mathrm{Im}}\nolimits}

\renewcommand {\phi}{\varphi}

\newcommand{\eps}{\varepsilon}

\begin{document}
  \title{
Microscopic model of Purcell enhancement in hyperbolic metamaterials}
%   \affiliation{a}
% \affiliation{b}
% \affiliation{b}
% \affiliation{c}

  \author{Alexander N. Poddubny,$^{1,2}$, Pavel A. Belov,$^{1,3}$,
Pavel Ginzburg,$^4$ Anatoly V. Zayats,$^4$ and Yuri S. Kivshar$^{1,5}$}

    \affiliation{$^{1}$National Research University for Information Technology, Mechanics and Optics
(ITMO), St.~Petersburg 197101, Russia\\
$^{2}$Ioffe Physical-Technical Institute of the Russian Academy of Science,
St.~Petersburg 194021, Russia\\
$^{3}$School of Electronic Engineering and Computer Science, Queen Mary University of London, London E1 4NS, UK\\ $^{4}$Department of Physics, King's College London, London WC2R 2LS, UK \\
$^{5}$Nonlinear Physics Center and Center for Ultrahigh-bandwidth Devices for
Optical Systems (CUDOS), Research School of Physics and Engineering, Australian National University,
Canberra ACT 0200, Australia}


\begin{abstract}
We study theoretically a dramatic enhancement of  spontaneous emission in metamaterials with the hyperbolic dispersion modeled as a cubic lattice of anisotropic resonant dipoles. We analyze the dependence of the Purcell factor on the source position in the lattice unit cell and demonstrate that the optimal emitter position to achieve large Purcell factors and Lamb shifts are in the local field maxima. We show that the calculated Green function has a characteristic cross-like shape, spatially modulated due to structure discreteness. Our basic microscopic theory provides fundamental insights into the rapidly developing field of hyperbolic metamaterials.
\end{abstract}

\maketitle

\section{Introduction}

Engineering light-matter interaction in nanostructured environment has recently been the focus of active studies~\cite{shalaev2011,Krishnamoorthy2012,khitrova2011,Welsch2006,Ivchenko2005,kavbamalas,maslovski2012}. In particular, the so-called hyperbolic metamaterials described by an uniaxial medium where the main components of dielectric and magnetic tensor have different signs, have attracted significant attention~\cite{Felsen,lindell2001,Smith2003}. Realizations of the regime of the hyperbolic medium
with negative components of the dielectric tensor have been reported for magnetized plasma~\cite{fisher1969}, graphite~\cite{Sun2011}, for metamaterials based on nanorod assemblies~\cite{Wurtz2008,narimanov2009b,simovski2012} and for layered metal-dielectric structures~\cite{elser2007b,orlov2011,Krishnamoorthy2012}. In this regime, light wavevectors at a given frequency fill a surface of a hyperbolic shape, so that the area of hyperbolic isofrequency surface, giving the photonic density of states, is infinite. As the result, the spontaneous emission rate becomes infinite in the case of ideal hyperbolic medium~\cite{narimanov2010,Krishnamoorthy2012}.

Experimental reports on the Purcell factor enhancement, describing the spontaneous emission rate modification, in hyperbolic metamaterials are already available~\cite{Noginov2010,Ni2011,Kim2012,Krishnamoorthy2012}. Theoretical studies for various models have been also performed~\cite{Cortes2012Arxiv}. It has been shown that the Purcell factor
should not actually diverge. It is rather determined by a cutoff in the wavevector
space, stemming from spatial inhomogeneity of the medium~\cite{narimanov2009,maslovski2011,Iorsh2012}, a finite  distance from the source to the medium~\cite{Xie2009,sipe2011}, nonlocality of dielectric response~\cite{Yan2012arXiv}, or finite size of the emitter~\cite{poddubny2011pra}.

The basic solid state model of either natural or artificial material  is a periodic lattice of unit cells. We adopt this model and consider hyperbolic material as the infinite cubic crystal of interacting resonant point dipoles. Similar models have been developed for various systems including lattices of quantum dots~\cite{Ivchenko2000}, optical atomic lattices~\cite{deutsch1995,lagendijk1996}, $\gamma$-ray resonant nuclear scattering~\cite{kagan1999} as well as the discrete-dipole-approximation of the light scattering theory~\cite{Purcell1973}. This general approach, despite certain approximations, has been successfully applied to the lattices of split-ring resonators~\cite{belov2005,belov2008b}.

\begin{figure}
\begin{center}
\includegraphics[width=0.5\linewidth]{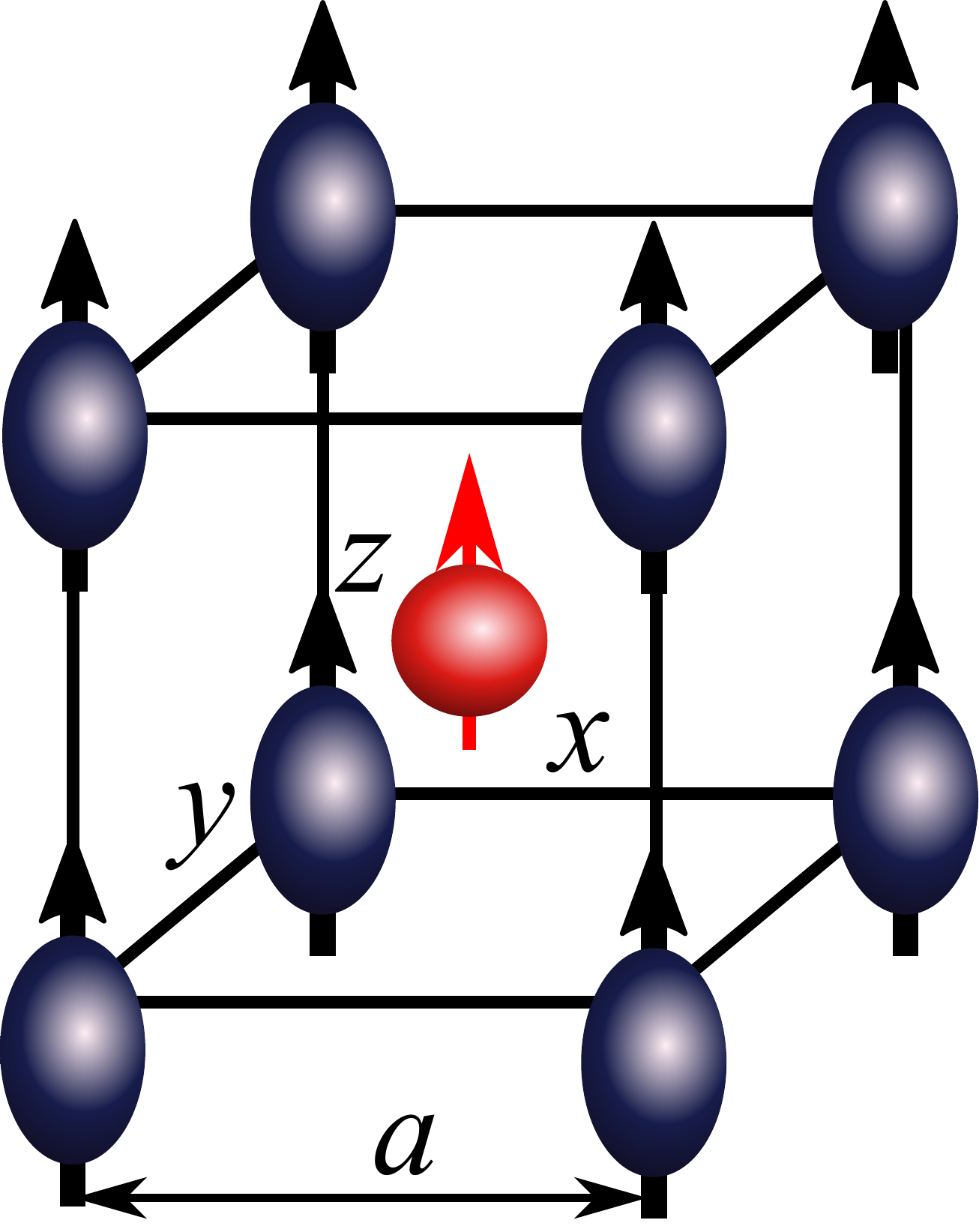}
\end{center}
 \caption{(Color online) Schematic illustration of the unit cell of the cubic
dipole lattice  with  embedded light source. }
\label{fig:1}
\end{figure}

In this paper, we study optical properties of the infinite cubic crystal of resonant interacting point dipoles polarizable only in one direction (see Fig.~\ref{fig:1}). This model allows us to reproduce the hyperbolic isofrequency surfaces of the uniaxial anisotropic metamaterials and accounts for the discrete character of  metamaterials. Within this microscopic model of hyperbolic metamaterial, we investigate the influence of emitter position within the unit cell of the metamaterial on its radiation properties.

The paper is organized as follows. Section~\ref{sec:model} outlines our theoretical model and approach. Calculated dispersion and lattice Green function are discussed in Sec.~\ref{sec:disp}.
Section~\ref{sec:purc} is devoted to the numerical and analytical results for the
Purcell factor and Lamb shift in metamaterials with hyperbolic dispersion.

% % % % % % % % % % % % % % % % % % % % % % % % %
% % % % % % % % % % % % % % % % % % % % % % % % % % % % % %

\section{Discrete dipole model}\label{sec:model}
% % % % % % % % % % % % % % % % % % % % % % % % % % % % % %
We consider an infinite periodic cubic lattice $\bm r_j$ of the point dipoles, characterized by
the period $a$, and embedded in vacuum. Our approach can be straightforwardly generalized to allow for
background dielectric constant $\varepsilon\ne 1$.
The emitter inside the lattice is modeled by a radiating dipole $\bm p_0$ is placed at
the point $\bm r_0$. Structure geometry is sketched on Fig.~\ref{fig:1}. The self-consistent electric field  satisfies the following equation
\begin{equation}\label{eq:E}
 \nabla\times\nabla\times \bm E-q^2\bm E=4\pi q^2\bm P\:,
\end{equation}
where $q=\omega/c$ is the wavevector at the frequency $\omega$.
The quantity $\bm P$ in Eq.~\eqref{eq:E} is
the net polarization of the lattice dipoles and the emitter:
\begin{equation}\label{eq:P}
 \bm P=\bm d_0\delta(\bm r-\bm r_0)+\sum\limits_{j}\bm
p_j\delta(\bm r-\bm r_j)\:.
\end{equation}
All the dipoles $\bm p_j$ are characterized by the identical polarizability tensor
$\hat\alpha$
\begin{equation}
 \bm p_j=\hat\alpha \bm E_{\rm ext}(\bm r_j)\:,
\end{equation}
%describing their response to the applied external electric field.
Our goal is to determine  the total electric field and polarizations, induced
in the structure by the radiating dipole $\bm d_0$. This procedure includes field expansion over the Bloch eigenmodes with wavevectors $\bm k$, for which Eqs.~\eqref{eq:E},\eqref{eq:P} are independent. The results in a coordinate space are obtained by inverse Fourier transformation.
In particular, the polarizations of lattice dipoles read
\begin{equation}\label{eq:pj}
\bm p_j=\int_{\rm (BZ)} \frac{V_0\rmd^3k}{(2\pi)^3}\e^{\rmi \bm k\bm r_j}
\hat\alpha\left[\hat 1-\hat C(\bm k)\hat\alpha\right]^{-1}\hat G_{0,\bm k}(-\bm r_0)\bm
d_0
\end{equation}
where $V_0=a^3$ is the unit cell volume, the integration takes place over the Brillouin zone
$|k_{m}|<\pi/a$, $m=x,y,z$, and $\hat 1$ is $3\times 3$ unity matrix.
 Radiating dipole position $\bm r_0$ enters Eq.~\eqref{eq:pj} and thus determines the efficiency of the lattice excitation.
The quantity $\hat C$ in Eq.~\eqref{eq:pj} is the tensor interaction constant of
the lattice, defined as~\cite{belov2005}
\begin{equation}\label{eq:C}
 \hat C(\bm k)=\lim_{\bm r\to 0} [\hat G_{0,\bm k}(\bm r)-\hat G_0(\bm
r)]+\frac{2\rmi q^3}{3}\hat 1\:,
\end{equation}
where $\hat G_{0,\bm k}$ is the Green function of the photon with Bloch vector
$\bm k$,
\begin{equation}\label{eq:G0k}
 \hat G_{0,\bm k}(\bm r)=\sum\limits_{j}\hat G_0(\bm r-\bm r_j)\e^{\rmi\bm k\bm
r_j}\:.
\end{equation}
and $\hat G_0$ is the free photon Green function
\begin{equation}\label{eq:G0}
 \hat G_0(\bm r)= \left[q^2+\nabla\nabla\right]\hat 1\frac{\e^{\rmi q r}}{r}\:.
\end{equation}
The infinite lattice sums \eqref{eq:G0k} may be found either by Ewald
summation~\cite{Korringa1947} or by a Floquet-type
summation~\cite{belov2005}. We have used the approach from Ref.~\onlinecite{belov2005}, since it is preferential for fast evaluation of the integral \eqref{eq:pj}.
Electric field in the structure is given by the sum of the waves emitted by all
the dipoles\:,
\begin{multline}\label{eq:E2}
 \bm E(\bm r)=\hat G_0(\bm r-\bm r_0)\bm p_0+\sum\limits_{j}\hat G_0(\bm r-\bm
r_j)\bm p_j\:.
\end{multline}
Eq.~\eqref{eq:E2}, by definition provides the Green function for the source embedded in the
dipole lattice.
Second term in Eq.~\eqref{eq:E2} is
given by Eq.~\eqref{eq:pj} where
 $\e^{\rmi \bm k\bm r_j}$ is replaced by $\hat G_{0,\bm k}(\bm r)$.
% % % % % % % % % % % % % % % % % % % % % % % % % % % %
\begin{figure}[t]
\includegraphics[width=\linewidth]{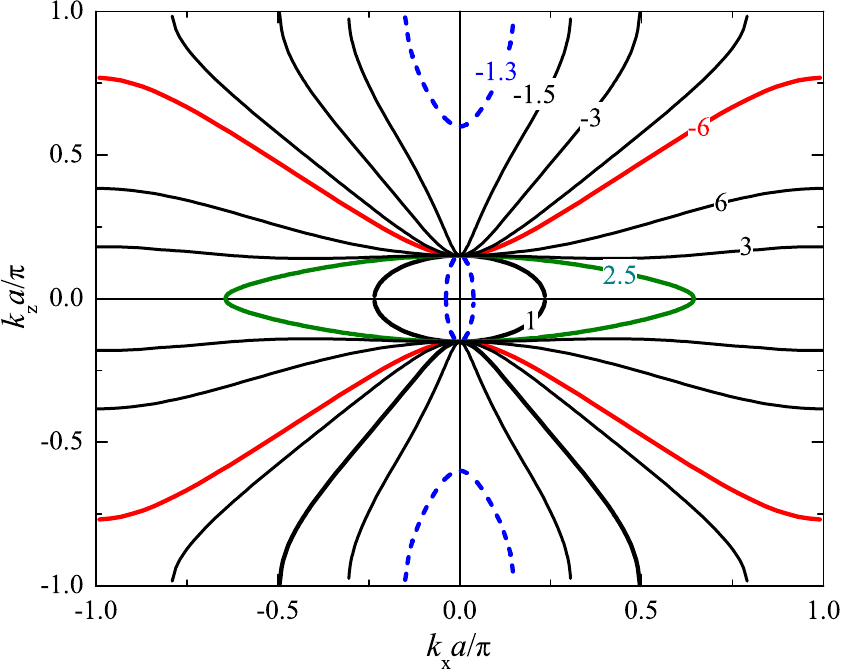}
 \caption{(Color online) Isofrequency curves in $xz$ plane,
calculated for different dipole polarizabilities $\alpha_{0,zz}$.
Normalized polarizability $4\pi \alpha_{0,zz}/a^3$ is indicated near each curve.
Calculated was performed at $qa=0.15 \pi$.
}\label{fig:disp}
\end{figure}

From now we restrict ourselves to the case of uniaxial dipoles, when
the only non-zero component of the tensor $\hat\alpha$ is $\alpha_{zz}$. We assume  that the the radiating dipole $\bm d_0$ is also directed along $z$ axis.
The TM-polarized Bloch eigenmodes of the system with given wavevector $\bm k$ are
found\cite{belov2005,belov2008b} from the poles of Eq.~\eqref{eq:pj}
\begin{equation}\label{eq:disp}
 \frac{1}\alpha_{zz}-C(\bm k)=0\:,
\end{equation}
where $C(\bm k)\equiv C_{zz}(\bm k)$.
Note, that Eq.~\eqref{eq:disp} is real for vanishing losses, because the
imaginary part of the interaction constant \eqref{eq:C} cancels out with the
radiative decay term in the polarizability:
\begin{equation}
 \frac{1}{\alpha_{zz}}=\frac{1}{\alpha_{0,zz}}-\frac{2\rmi q^3}{3}\hat 1\:.
\end{equation}
Here $\alpha_{0,zz}$ is the so-called bare dipole polarizability, calculated
neglecting radiative decay~\cite{lagendijk_review}.
Effective medium approximation for the solutions of Eq.~\eqref{eq:disp} are the
extraordinary TM-polarized modes, with the dispersion given by~\cite{landau08}
\begin{equation}\label{eq:disp_eff}
 q^2=\frac{k_x^2+k_y^2}{\varepsilon_{zz}}+k_z^2\:.
\end{equation}
Here $\varepsilon_{zz}$ is the Maxwell-Garnett effective dielectric constant of
the hyperbolic medium
\begin{equation}\label{eq:MG}
 \varepsilon_{zz}=1+\frac{1}{V/(4\pi\alpha_{0,zz})-1/3}\:,
\end{equation}
in the same approximation $\varepsilon_{xx}=\varepsilon_{yy}=1$.

% % % % % % % % % % % % % % % % % % % % % % % % % % % % % %
\section{Dispersion and Green function}\label{sec:disp}
% % % % % % % % % % % % % % % % % % % % % % % % % % % % % % %
% % % % % % % % % % % % % % % % % % % % % % % % % % % % % % % % % %
\begin{figure}[t]
\centering\includegraphics[width=0.8\linewidth]{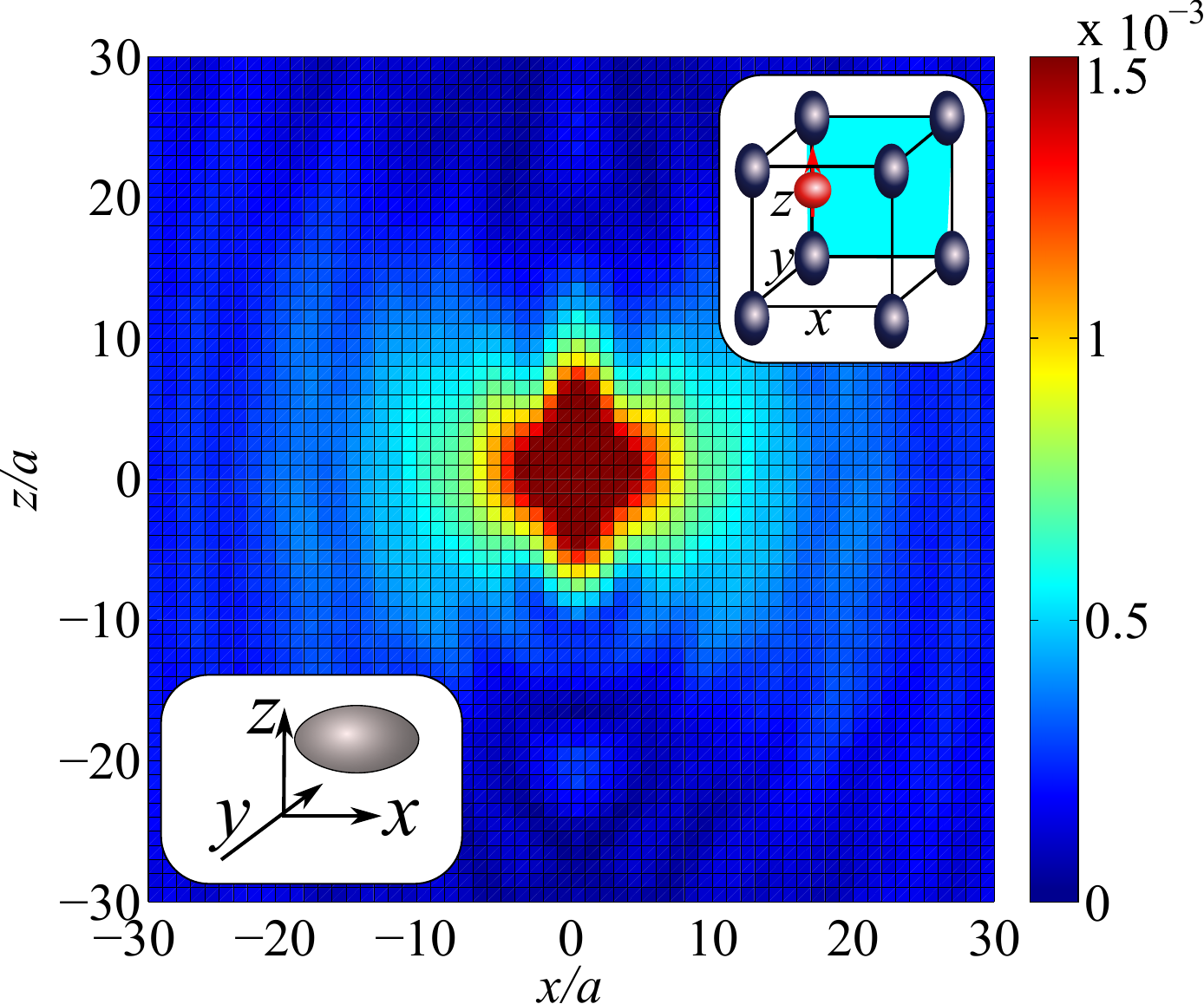}
 \caption{(Color online) Spatial distribution of the dipole moments
$|p_z(\bm r)|/p_0$ in the elliptic regime with $\alpha_{0,zz}=a^3/(4\pi)$.
  Insets schematically illustrate the geometry and the isofrequency surfaces in wavevector space.
Calculation was performed  at $qa=0.15\pi$ and $\bm r_0=0.5a\hat{\bm z}$.
}\label{fig:spat_hypell}
\end{figure}
% % % % % % % % % % % % % % % % % % % % % % % % % % % %
In this Section we first discuss the dispersion of the Bloch waves in the dipole lattice (Sec.~\ref{sec:disp0}) and then investigate in detail the emission pattern of the  dipole embedded in the lattice (Sec.~\ref{sec:numGreen}).
% % % % % % % % % % % % % % % % % % % % % % % % % % % % % % %

% % % % % % % % % % % % % % % % % % % % % % % % % % % % % % % % % %
\begin{figure}[t]
\centering\includegraphics[width=0.8\linewidth]{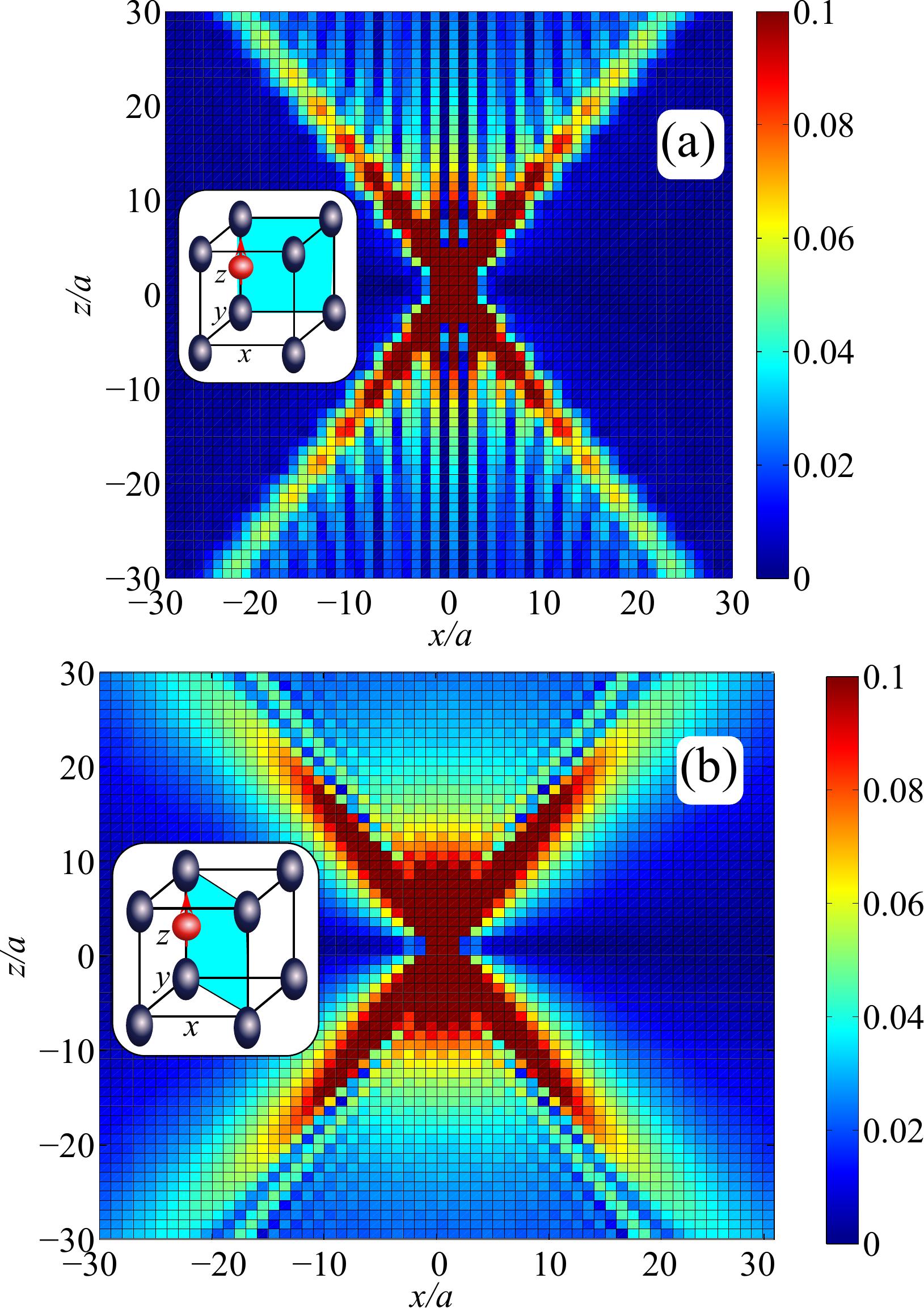}
 \caption{(Color online) Spatial distribution of the dipole moments
$|p_z(\bm r)|/p_0$ in the hyperbolic regime, excited by the point emitter.
Panels (a) and (b) show the distribution in the planes $y=0$ and $x=y$, respectively. Insets schematically illustrate the geometry and the isofrequency surfaces in wavevector space.
  Calculation was performed at $\alpha_{0,zz}=-6a^3/(4\pi)$ and the same other parameters as Fig.~\ref{fig:spat_hypell}\:.
}\label{fig:spat_hyp}
\end{figure}

% % % % % % % % % % % % % % % % % % % % % % % % % % % % % % % % % %
\begin{figure}[t]
\centering\includegraphics[width=0.8\linewidth]{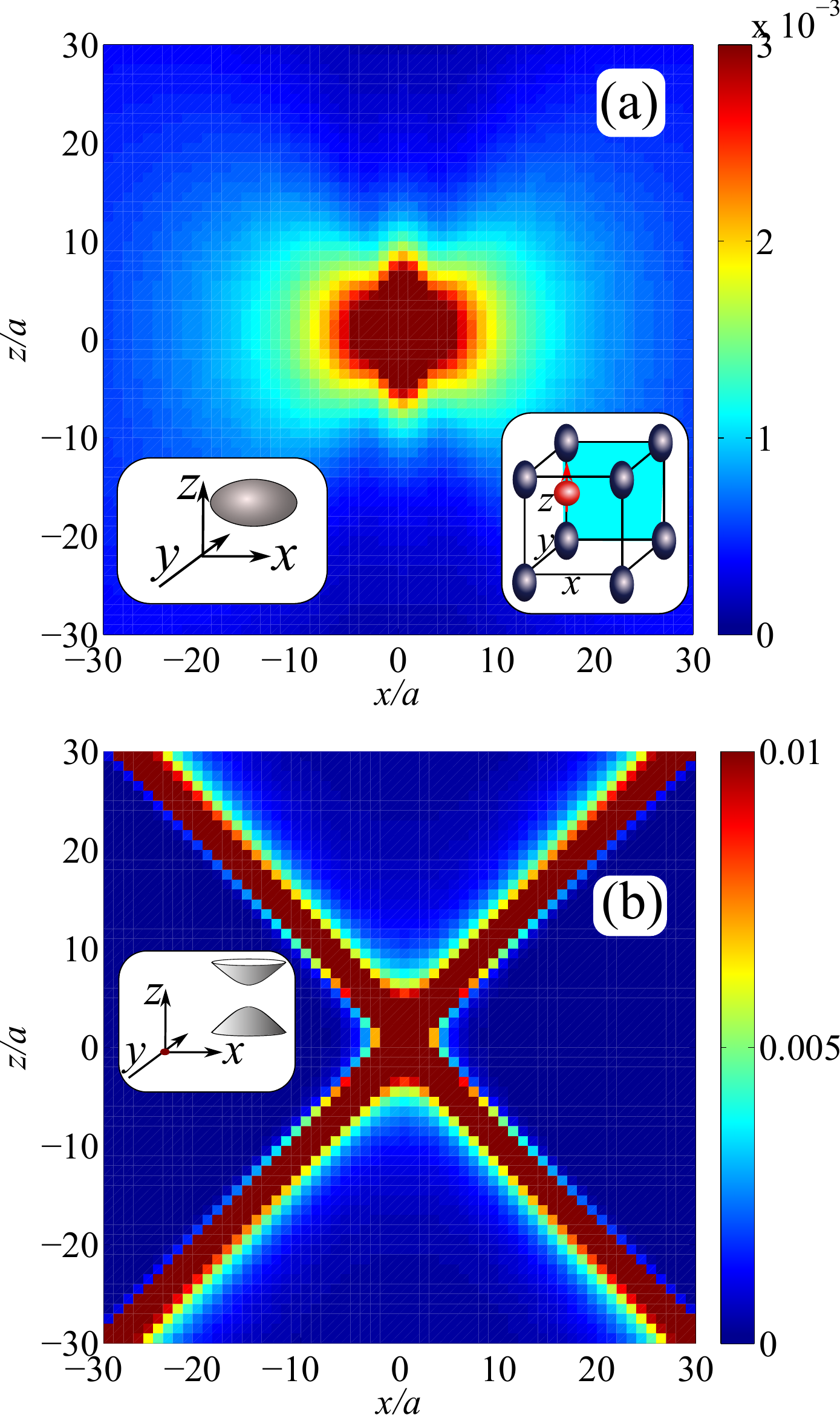}
 \caption{(Color online) Spatial distribution of the polarization $P_{{\rm eff},z}(\bm r)/(p_0a^3)$ induce by the point source in (a) effective elliptic medium with $\eps_{zz}=2.5$ and (b) effective hyperbolic medium with $\eps_{zz}=-1$. Insets schematically illustrate the geometry and the isofrequency surfaces in wavevector space. Polarization is evaluated at the discrete lattice sites $\bm r_j$ in the $xz$ plane.
Other calculation parameters are the same as for Fig.~\ref{fig:spat_hypell}.
}\label{fig:spat_hypell_eff}
\end{figure}
% % % % % % % % % % % % % % % % % % % % % % % % % % % % % % % % % %
\begin{figure}
\includegraphics[width=\linewidth]{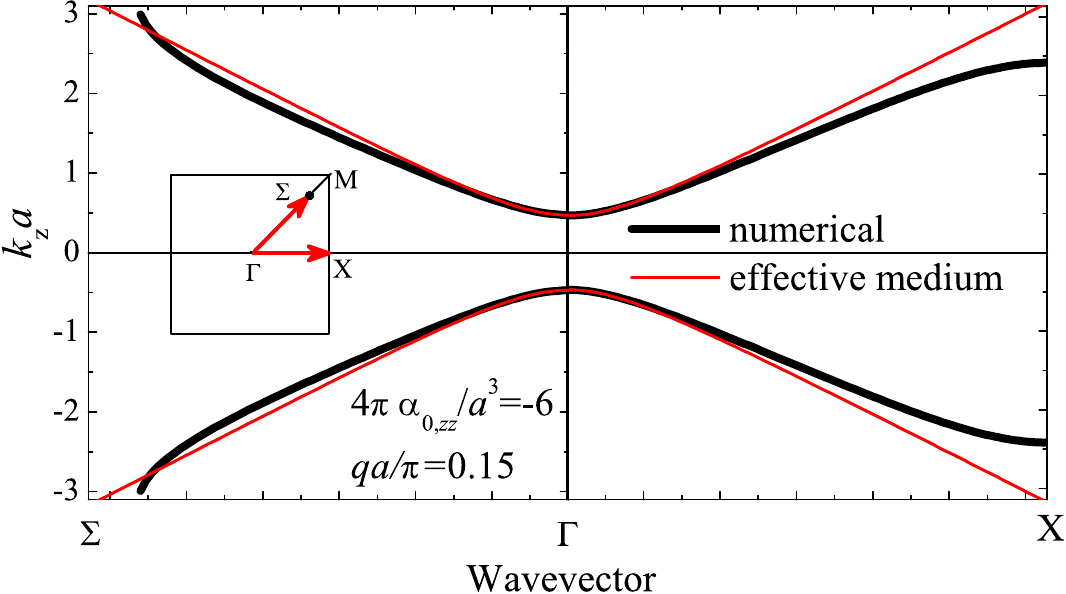}
 \caption{Isofrequency curves of the dipole lattice in the hyperbolic regime.
Solid and thin lines correspond to numerical calculation and effective medium
approximation Eq.~\eqref{eq:disp_eff}. Other parameters are the same as for Fig.~\ref{fig:spat_hyp}.
The inset schematically indicates the Brillouin zone of the square lattice,
point $\Sigma$ corresponds to $k_x=k_y=\pi/(\sqrt{2}a)$\:.
}\label{fig:disp_hyp}
\end{figure}
% % % % % % % % % % % % % % % % % % % % % % % % %
\subsection{Isofrequency curves}\label{sec:disp0}
% % % % % % % % % % % % % % % % % % % % % % % % %
Isofrequency curves in the $(k_z,k_x)$ plane,  found from Eq.~\eqref{eq:disp} for different polarizabilities $\alpha_{0,zz}$, are shown on Fig.~\ref{fig:disp}. Depending on the polarizability, dispersion curves are either elliptic or hyperbolic, in agreement with Eq.~\eqref{eq:MG}. The curves are generally well described by the effective medium approximation \eqref{eq:disp_eff}. However, an intermediate ``mixed'' regime is possible for $\alpha_{0,zz}\approx -1.3 a^3/(4\pi)$ (blue dashed curve), when two Bloch modes with hyperbolic and elliptic dispersion coexist in the structure. Such isofrequency curves can not be described by the Maxwell-Garnett theory Eqs.~\eqref{eq:disp_eff},\eqref{eq:MG}, which predicts only one TM mode for given frequency. They were  analyzed in Ref.~\onlinecite{belov2005} in more details and can be obtained in the effective medium model when nonlocal effects are taken into account~\cite{pollard2009,belov2003}.

% % % % % % % % % % % % % % % % % % % % % % % % % % % % % % %
\subsection{Green function}\label{sec:numGreen}
% % % % % % % % % % % % % % % % % % % % % % % % % % % % % % %
Here we will focus on the spatial distribution \eqref{eq:pj} of the dipole
moments $|p(\bm r_j)|$ in the discrete lattice under the point dipole excitation.
Results of calculation for the dipole polarizabilities $\alpha_{0,zz}=a^3/(4\pi)$ and
$\alpha_{0,zz}=-6a^3/(4\pi)$, corresponding to elliptic and hyperbolic regimes, are shown on Fig.~\ref{fig:spat_hypell} and Fig.~\ref{fig:spat_hyp}, respectively. Calculation demonstrates that the distribution is qualitatively different in hyperbolic regime: the pattern is strongly anisotropic and has characteristic cross-like shape, see Fig.~\ref{fig:spat_hyp}. Moreover, in hyperbolic case the pattern depends on the azimuthal direction: it has distinct vertical ripples in the plane $y=0$ (Fig.~\ref{fig:spat_hypell}a), which are absent in the plane $y=x$ (Fig.~\ref{fig:spat_hypell}b).

To understand these results it is instructive to compare them with  Green function in the
effective medium approximation, see Fig.~\ref{fig:spat_hypell_eff}. This approximation allows one to obtain the solution  in a closed form~\cite{Felsen,savchenko2005}.
 In  the case  of a vertical orientation  of the radiating dipole, $\bm p_0 \parallel z$, Green function reads
\begin{multline}
 \bm E_{{\rm eff}}(\bm r)=(q^2+\nabla\nabla)p_0\hat{\bm z}\frac{\e^{\rmi qR}}{R},\\ R=\sqrt{\eps_{zz}(x-x_0)^2+\eps_{zz}(y-y_0)^2+(z-z_0)^2}\:.\label{eq:Green_hom}
\end{multline}
This is generalization of Eq.~\eqref{eq:G0} in the case of uniaxial medium.
The relation between electric field and polarization in the effective medium model is local,
\begin{equation}\label{eq:PGreen_hom}
 4\pi \bm P_{\rm eff}=(\eps_{\rm eff}-1)\bm E_{\rm eff}\:.
\end{equation}
It should be stressed that the  effective medium approximation is not applicable on the spatial scales smaller than the lattice constant $a$. Consequently,
the problem of point radiating dipole in discrete structure can not be reduced to the effective medium one. The effects of the radiating dipole position within the unit cell are also beyond the effective medium approximation. Thus, the results of two models, discrete and effective, may be compared only qualitatively.

 In order to clarify the ambiguity we have evaluated on Fig.~\ref{fig:spat_hypell_eff} the polarization \eqref{eq:PGreen_hom} at the discrete set of square lattice points $\bm r_j$ and the radiating dipole is located at the point $\bm r_0=0.5a\hat{\bm z}$, see the inset of panel (a).  Fig.~\ref{fig:spat_hypell_eff}a and Fig.~\ref{fig:spat_hypell_eff}b
show the spatial distribution of the polarization Eq.~\eqref{eq:PGreen_hom}  for the values effective dielectric constants off $\eps_{zz}=2.5$ and $\eps_{zz}=-1$, corresponding to Fig.~\ref{fig:spat_hypell} and Fig.~\ref{fig:spat_hyp}. In the case of the material with elliptic dispersion the emission pattern is qualitatively the same as for the isotropic medium. The near field is concentrated at the emitter origin, $\bm r=\bm r_0$, while the far-field is emitted perpendicularly to the dipole axis. The pattern changes dramatically  in the hyperbolic case (Fig.~\ref{fig:spat_hypell_eff}b). The distribution has a distinct cross-like shape, typical for hyperbolic medium~\cite{fisher1969,Felsen}. In the elliptic case, the only field singularity is that at the origin $ R=|\bm r-\bm r_0|=0$. In the hyperbolic medium this singular point becomes a conical surface, where the field intensity is concentrated. Radiated waves are propagating within the cone $R^2>0$, and are evanescent outside this cone. Energy flow directions are normal to the isofrequency surfaces, so such cone in $\bm r$-space is a direct counterpart  of the hyperbolic dispersion curves in  $\bm k$-space.

Comparing numerical and effective medium results, Fig.~\ref{fig:spat_hypell} and
 Fig.~\ref{fig:spat_hypell}a, we see that in the elliptic  case the Green function  is qualitatively the same as in the effective medium approximation. Weak spatial modulation of the dipole moments $|p(\bm r_j)|$, seen on  Fig.~\ref{fig:spat_hypell}, is related to the deviations from the effective medium theory Eq.~\eqref{eq:Green_hom}, which, as mentioned above, is not completely valid for the point excitation.
  Distinct cross-like distribution of  Fig.~\ref{fig:spat_hyp}a is a fingerprint of the hyperbolic regime, similar to the  effective medium approximation of Fig.~\ref{fig:spat_hypell_eff}b.
Comparing Fig.~\ref{fig:spat_hypell_eff}b and Fig.~\ref{fig:spat_hyp}a, we  see,  that in the discrete case the singularity in the effective medium solution \eqref{eq:Green_hom} at the conical surface $R=0$ is smeared out  and even vanishes at large enough distances, where the effective approximation also breaks down. This is qualitatively explained by the presence of the wavevector cutoff $\sim \pi/a$.
The second striking difference between Fig.~\ref{fig:spat_hyp} and its effective medium counterpart Fig.~\ref{fig:spat_hypell_eff}b is the strong spatial modulation of the distribution in the $y=0$ plane, manifested as  vertical ripples. Such modulation is obviously beyond the effective medium approximation of Fig.~\ref{fig:spat_hypell_eff}b. In particular, the ripples on Fig.~\ref{fig:spat_hyp}(a) turn out to be the result of the interference of the Bloch waves with wavectors $k_x=\pm \pi/a$, corresponding to the boundaries of the Brillouin zone.
To check this hypothesis we have plotted  on Fig.~\ref{fig:disp_hyp} the isofrequency curves in $\Gamma-$X and $\Gamma-$M directions.  Interference pattern in the planes $y=0$ and $y=x$ should depend on the dispersion along $\Gamma-$X and $\Gamma-$M, respectively.
 Since $dk_z/dk_x=0$ at $k_x=\pi/a$ (right panel of Fig.~\ref{fig:disp_hyp}), there is a singularity in the  density of states propagating along $x$ direction, promoting the vertical ripples.
This singularity is absent for the $\Gamma-$M direction, where  the behavior of the isofrequency curves at the Brillouin zone edge  is different.
 Panels (a) and (b)  of Fig.~\ref{fig:spat_hyp} present the spatial distribution \eqref{eq:pj} of the dipole  moments $|p(\bm r_j)|$ in the planes $y=0$
 and $y=x$, respectively.
We see that the spatial modulation in the plane $y=x$  is absent, cf.  Fig.~\ref{fig:spat_hyp}a and
Fig.~\ref{fig:spat_hyp}b, which corroborates our explanation.
The discovered effect can be thought of as the manifestation of the Van Hove band edge singularity~\cite{KittelIntro} in the Green function~\cite{Swendsen1972}.
% % % % % % % % % % % % % % % % % % % % % % % % % % % % % % % % % %
\begin{figure}[t]
\centering\includegraphics[width=0.8\linewidth]{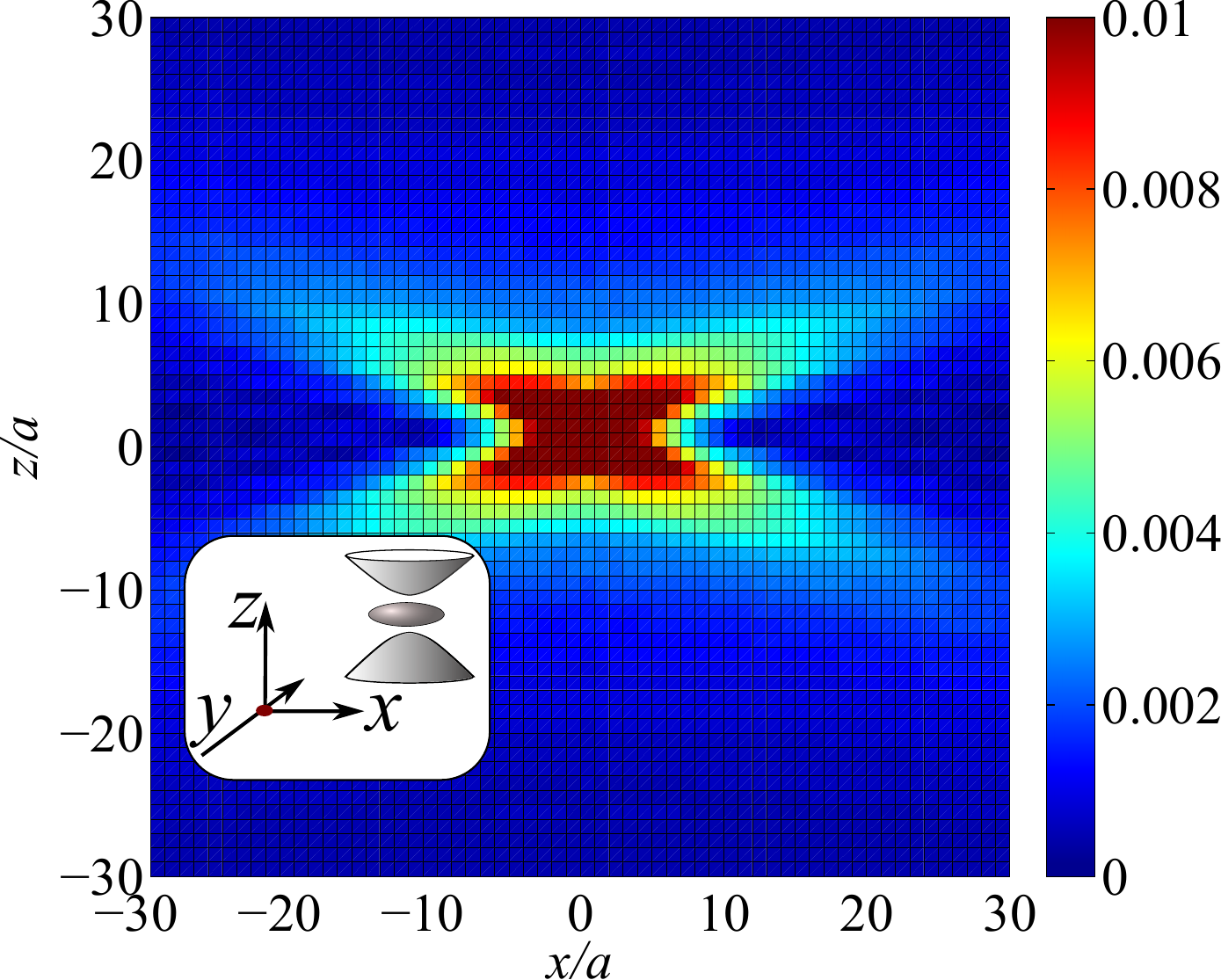}
 \caption{(Color online) Spatial distribution of the dipole moments
$|p_z(\bm r)|/p_0$ in the mixed hyperbolic-elliptic regime with $\alpha_{0,zz}=-1.3a^3/(4\pi)$.
  Insets schematically illustrate the geometry and the isofrequency surfaces in wavevector space.
Other calculation parameters are the same as for Fig.~\ref{fig:spat_hypell_eff}.
}\label{fig:spat_mixed}
\end{figure}

Discrete Green function  calculated for the dipole polarizability
$\alpha_{0,zz}=-1.3a^3/(4\pi)$, corresponding to mixed elliptic-hyperbolic regime, is shown on Fig.~\ref{fig:spat_mixed}.  In this case the ripples  are absent, because the isofrequency curves do not reach the Brillouin zone boundary $k_x=\pm \pi/a$, see dashed curve on Fig.~\ref{fig:disp}. Far-field emission along $x$ direction is possible due to the modes with  elliptic dispersion,  providing weak background to the field of the hyperbolic modes.
% Fig.~\ref{fig:disp_hyp} presents the comparison of the exact isofrequency curves
% $k_z(k_x,k_y)$, numerically found from Eq.~\eqref{eq:disp} with the effective
% medium approximation Eq.~\eqref{eq:disp_eff}.
% Both curves have a distinct hyperbolic shape.
% Approximate  Eq.~\eqref{eq:disp_eff} well describes the numerical solution,
% excepting the narrow ranges near the edges of the Brillouin zone. No additional
% modes, discovered in Ref.~\onlinecite{belov2005}, are manifested for the chosen
% parameters, when $\varepsilon_{zz}=-1$.
%

% % % % % % % % % % % % % % % % % % % % % % % % % % % % % % %
% \subsection{Green function in effective medium approximation}\label{sec:eff}
% % % % % % % % % % % % % % % % % % % % % % % % % % % % % % %

% % % % % % % % % % % % % % % % % % % % % % % % % % % %
\section{Purcell factor}\label{sec:purc}

Here we investigate the role of the discreetness on the Purcell factor determining the characteristics of the spontaneous emission of the radiating dipole inside the material.
The Purcell factor $f$ and the Lamb shift $l$ for the radiating dipole  can be found from the Green function  \eqref{eq:E2}, evaluated at the dipole
origin~\cite{tomas1999,Ivchenko2005,Welsch2006}, see also Ref.~\onlinecite{kivshar2004}:
\begin{equation}\label{eq:fl0}
 f+\rmi l=1+\frac{3\rmi E_z(\bm r_0)} {2q^3 p_0} \:.
\end{equation}
Here the dimensionless Lamb shift $l$ is formally understood as a radiative
correction to the resonance frequency of the radiating two-level system, normalized
to its free space decay rate.
Gathering Eqs.~\eqref{eq:fl0},\eqref{eq:E2},\eqref{eq:pj} together, we find the result in a compact form
\begin{equation}\label{eq:f}
 f+\rmi l=\frac{3\rmi}{2q^3}\int\limits_{(BZ)}\frac{V_0\rmd^3 k }{(2\pi)^3}
\frac{|G_{\bm k,zz}(\bm r_0)|^2}{1/\alpha_{zz}- C(\bm k)-\rmi 0}\:.
\end{equation}
The frequency $\omega$, entering the wavevector $q$ in Eq.~\eqref{eq:f},
is determined by the transition energy of the radiating dipole $\bm d_0$.
It is clear from the structure of Eq.~\eqref{eq:f}, that the Purcell factor is determined by the pole contribution, corresponding to emission of photons with the dispersion given by Eq.~\eqref{eq:disp}. We note, that the first term in right-hand-side of Eq.~\eqref{eq:fl0} has canceled out in Eq.~\eqref{eq:f} with the pole contribution in the free space Green function $G_{\bm k,zz}(\bm r_0)$ at $q=k$. We  stress, that despite the classical formulation of the problem, the results for the emission rate and photon Green function may be equivalently obtained by  the  quantum-mechanical calculation, either using the Fermi Golden rule~\cite{Ivchenko2005}  or the local quantization framework~\cite{Welsch2006}.

\begin{figure}
\includegraphics[width=\linewidth]{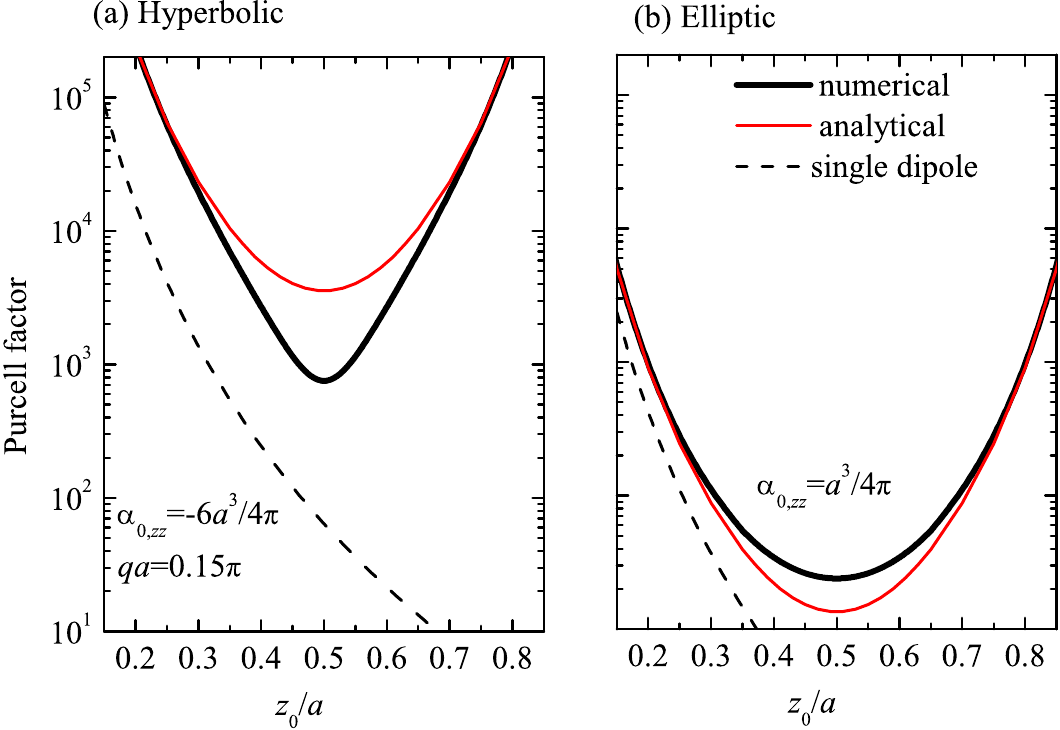}
 \caption{(Color online) Purcell factor in the (a) hyperbolic
and (b) elliptic regime as function of the source coordinate $z_0$ for $x_0=y_0=0$.
Thick solid black,  thin solid red, and dashed black curves correspond to
numerical calculations, the analytical results of Eq.~\eqref{eq:f_an}(panel a), Eq.~\eqref{eq:f_an2} (panel b), and to a single dipole with corresponding polarizability (Eq.~\eqref{eq:f_2}), respectively. Dashed curves are results \eqref{eq:f1} for source  near single dipole at $\bm r=0$.
Other parameters  as the same as for Fig.~\ref{fig:spat_hyp}.
}\label{fig:hyp_ell}
\end{figure}

\begin{figure}[b]
\includegraphics[width=\linewidth]{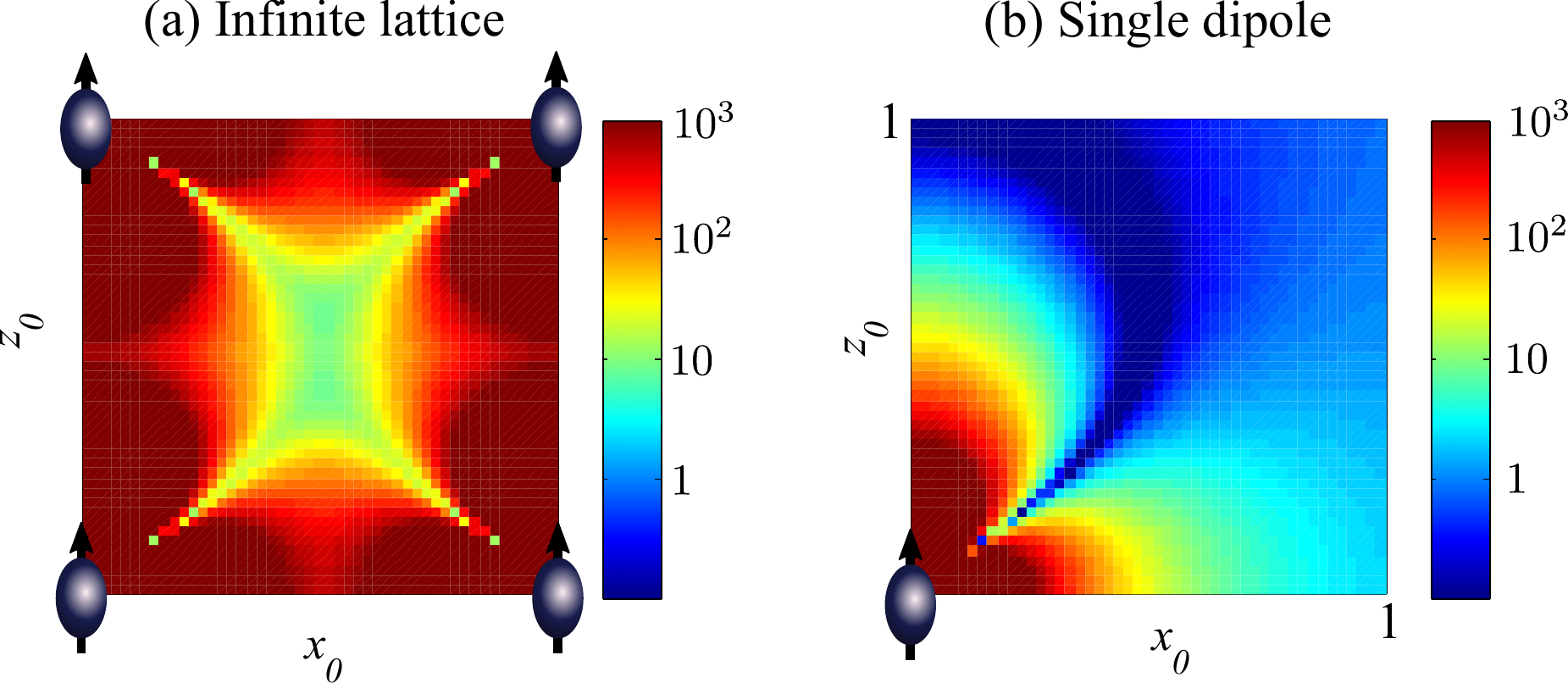}
\caption{(Color online)
(a) Purcell factor in hyperbolic medium as function of the  source position in  the unit cell. (b) Calculation in   single-dipole approximation Eq.~\eqref{eq:f_2}.
Calculation was carried out at $y_0=0$ and the same other parameters as for Fig.~\ref{fig:spat_hyp}. Radiating dipole coordinates change within the square $0\le x_0\le 1$, $0\le z_0\le 1$.
Colors correspond to logarithmic scale, identical for both panels.
}\label{fig:move2d}
\end{figure}

\begin{figure}[t]
\includegraphics[width=\linewidth]{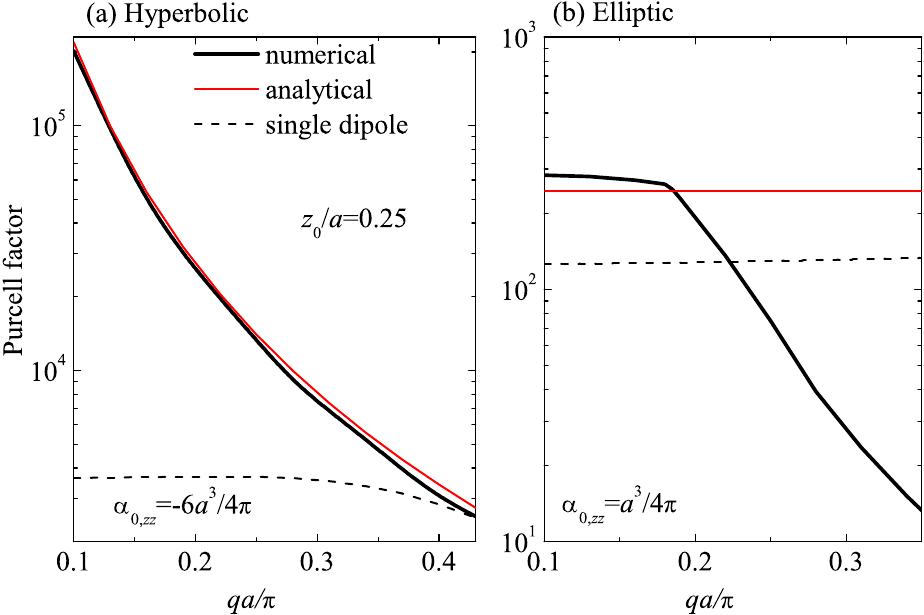}
 \caption{(Color online) Purcell factor in  (a) hyperbolic
and (b) elliptic regime as function of the frequency $qa$ for $\bm r_0=0.25a\hat{\bm z}$.
Notation and other parameters  as the same as for Fig.~\ref{fig:hyp_ell}.
}\label{fig:hyp_ell_q}
\end{figure}

Numerical  results for the dependence of the Purcell factor on the radiating dipole position within the unit cell of the structure are  presented in Figs.~\ref{fig:hyp_ell},
Fig.~\ref{fig:move2d}. Fig.~\ref{fig:hyp_ell_q} shows the frequency dependence of the Purcell factor.
Figures demonstrate that the Purcell factor is much larger in hyperbolic regime than in the elliptic one. It is very  sensitive to the dipole position and strongly increases when the dipole approaches the lattice nodes. Before discussing these results in more details it is instructive to compared them with the analytical theory.

% % % % % % % % % % % % % % % % % % % % % % % % % % % % % % % % % %
% % % % % % % % % % % % % % % % % % % % % % % % % % % %
% % % % % % % % % % % % % % % % % % % % % % % % % % % %

% % % % % % % % % % % % % % % % % % % % % % % % % % % %
% % % % % % % % % % % % % % % % % % % % % % % % % % % %

\subsection{Analytical results}

Here we focus on the Purcell factor in the quasi-static
limit $q\ll \pi/a$.
Eq.~\eqref{eq:f} can be then reduced to
\begin{multline}\label{eq:f1}
 f=
\frac{3}{2q^3}\frac{V_0|G_{zz,\rm stat}(\bm r_0)|^2}{(2\pi)^2}\int dk_x dk_y
\left|\frac{d C(\bm k)}{dk_z}\right|^{-1}_{k_{z}(k_x,k_y)}\:,
\end{multline}
where the interaction constant in the effective medium approximation reads
\cite{belov2008b}
\begin{equation}\label{eq:C1}
 C(\bm k)=\frac{4\pi }{V_0}\frac{q^2-k_z^2}{k^2-q^2}+\frac{4\pi}{3 V_0
}+\frac{2\rmi q^3}{3}\:.
\end{equation}
The integral over $k_z$ in \eqref{eq:f} is determined by the residues at the
wavevectors $\pm k_{z}(k_x,k_y)$, being the solutions of Eq.~\eqref{eq:disp} at
given frequency. Integration over $k_x$ and $k_y$ in Eq.~\eqref{eq:f1} is restricted
to those vectors within the two-dimensional Brillouin zone, for which such
solution exists.
The quantity
$G_{zz,\rm stat}(\bm r_0)$ in \eqref{eq:f1} is the quasistatic approximation of the Green function
\eqref{eq:G0k}: $G_{zz,\rm stat}(\bm r_0)\equiv G_{\bm k,zz}(\bm r_0)|_{\bm k=0,q=0}$. The value of $G_{zz,\rm stat}$  is determined by the near field
of the lattice dipoles, closest to the radiating one.
Maximum Purcell factor can be then expected when the source is
located on the vertical edge of the elementary cell, i.e. $x_0=y_0=0,z_0\ne 0.$
In this case $G_{zz,\rm stat}(z_0)$ can be reduced to
\begin{equation}
 G_{zz,\rm
stat}(z_0)\approx \frac{2}{z_0^3}+\frac{2}{(a-z_0)^3}
% -\frac{2}{(2a-z_0)^3}-\frac{2}{
% (a+z_0)^3}\:
\end{equation}
and grows dramatically when the emitter approaches the lattice node.
Evaluating the derivative in Eq.~\eqref{eq:f1}
by means of Eq.~\eqref{eq:C1} and performing the integration, we obtain the
analytical result for the Purcell factor
\begin{equation}\label{eq:f_an}
 f_{\rm hyp}=\frac{(\varepsilon-1)^2}{32\pi^2}\left(\frac{k_{z, \rm max}}{
q}\right)^3|V_0G_{zz,\rm stat}(z)|^2\:
\end{equation}
in the hyperbolic medium.
Here $k_{z,\rm max}\gg q$ is the cutoff for the wavevector $k_z$, existing due to the finite extent of the Brillouin zone.
Its value depends on the effective dielectric constant,
\begin{equation}
k_{z,\rm max}\approx \begin{cases}
\dfrac{\pi}{a},
           &-1\le \varepsilon\le 0\\
\dfrac{\pi}{a\sqrt{|\varepsilon_{zz}|}},
                                     &\varepsilon_{zz}\le -1\:.
 \end{cases}
\end{equation}
Thus, Eq.~\eqref{eq:f_an} provides a compact analytical result for the Purcell
factor in the lossless hyperbolic medium.
Its general structure can be understood as follows: the factor $(k_{z,\rm
max}/q)^3\sim 1/(qa)^3$ describes the enhancement of the photonic density of
states as compared to the vacuum. The second factor $|V_0G_{zz,\rm
stat}(z)|^2$ reflects the coordinate dependence of the Purcell factor, governed
by the near-field of the neighboring dipoles.
Near the lattice nodes  Eq.~\eqref{eq:f_an} can be simplified to
\begin{equation}\label{eq:f_div}
 f_{\rm hyp}(q,z\to 0)\approx \frac{\pi(\eps-1)^2a^3}{8q^3|z|^6}\:,
\end{equation}
where we assumed that $|\eps|\le 1$.

Similar calculation can be also performed in the elliptic case, when $\eps_{zz}>0$. It should be noted, that in the effective medium approximation, the Purcell factor for the axial dipole in the elliptic medium is unity, independently of the value of $\eps_{zz}$~\cite{poddubny2011pra}. Local-field corrections can still promote high decay rate. The answer reads
\begin{equation}\label{eq:f_an2}
 f_{\rm ell}=|G_{{\rm stat},zz}|^2 \left|\frac{V_0(\eps -1)}{4\pi}\right|^2\:.
\end{equation}
This expression depends on the local field intensity, similarly Eq.~\eqref{eq:f_an}, but is smaller by the factor
\begin{equation}\label{eq:gain}
 \frac{f_{\rm hyp}}{f_{\rm ell}}=\frac{k_{z,\rm max}^3}{2q^3}\:,
\end{equation}
 since the density of states in elliptic medium is smaller. In order to distinguish between the local field effects and the collective effects due to density of states enhancement in the medium it instructive to analyze also the Purcell factor for a source located in vacuum near {\it
single } dipole in the point $\bm r=0$. The result reads~\cite{Novotny2006}
\begin{equation}\label{eq:f_2}
 f_1=1+\frac{3}{2q^3}\Im [\alpha_{zz} G^2_{0,zz}(\bm r_0)]\:,
\end{equation}
here the second term is the field of the emitter, reflected from the dipole. In
the quasistatic limit $q\to 0$ Eq.~\eqref{eq:f_2} reduces to
\begin{equation}\label{eq:f_2b}
 f_1=\left(1+\frac{\alpha_{0,zz}}{|z^3|}\right)^2\:.
\end{equation}
Both Eqs.~\eqref{eq:f_div} and \eqref{eq:f_2b} demonstrate divergency when $z$
tends to zero. However, their dependence on the wavevector $q$ is quite
different: Eq.~\eqref{eq:f_div} diverges as $1/q^3$ at small $q$, while
Eq.~\eqref{eq:f_2b} does not depend on $q$ at all. This divergency is a
characteristic effect of  photonic density of states enhancement in the
hyperbolic medium~\cite{narimanov2010,poddubny2011pra,maslovski2011}.

% % % % % % % % % % % % % % % % % % % % % % % %
\subsection{Numerical results}
% % % % % % % % % % % % % % % % % % % % % % % %

Now we discuss the calculated dependence of the Purcell factor on
the source position  and on the transition frequency $\omega=cq$, shown on Fig.~\ref{fig:hyp_ell},
Fig.~\ref{fig:move2d} and Fig.~\ref{fig:hyp_ell_q}.
The
calculation confirms the singular behavior of the Purcell factor in the hyperbolic case when the source
approaches the lattice nodes (solid curve on Fig.~\ref{fig:hyp_ell}a). The singularity
is excellently described by analytical Eq.~\eqref{eq:f_an} (thin red curve). The interaction of the emitter with a single dipole  (Eq.~\eqref{eq:f_2}) provides substantially smaller enhancements (dashed black curve), although it also diverges at $z=0$ and $z=a$. Additional comparison of the exact calculation in medium with the result for a single dipole is presented by the Purcell factor dependence on the two coordinates $x$ and $z$ in the unit cell, shown on Fig.~\ref{fig:move2d}. Eq.~\eqref{eq:f_2}, taking into account only single lattice dipole at the  point $\bm r=0$, satisfactory reproduces the Purcell factor pattern near this point.
Corresponding two-dimensional plot of the Purcell factor  in the quadrant $0\le x,y\le 1$  is shown in Fig.~\ref{fig:move2d}(b). Comparing two panels of Fig.~\ref{fig:move2d} we see, that near the corner $\bm r=0$ the  angular dependence is approximately given by $(3\cos^2\theta-1)^2$, where $\theta$ is a polar angle. Still, single dipole model with corresponding polarizability $\alpha_{0,zz}=-6a^3/(4\pi)$ considerably underestimates the absolute values of the Purcell factor. The satiation is different in the elliptic case, where all three approaches, namely numerical calculation according to Eq.~\eqref{eq:f}, single dipole model \eqref{eq:f_2} with $\alpha_{0,zz}=a^3/(4\pi)$ and analytical model \eqref{eq:f_an2} provide similar results, see  Fig.~\ref{fig:hyp_ell}b.

\begin{figure}[t]
\begin{center}
\includegraphics[width=0.8\linewidth]{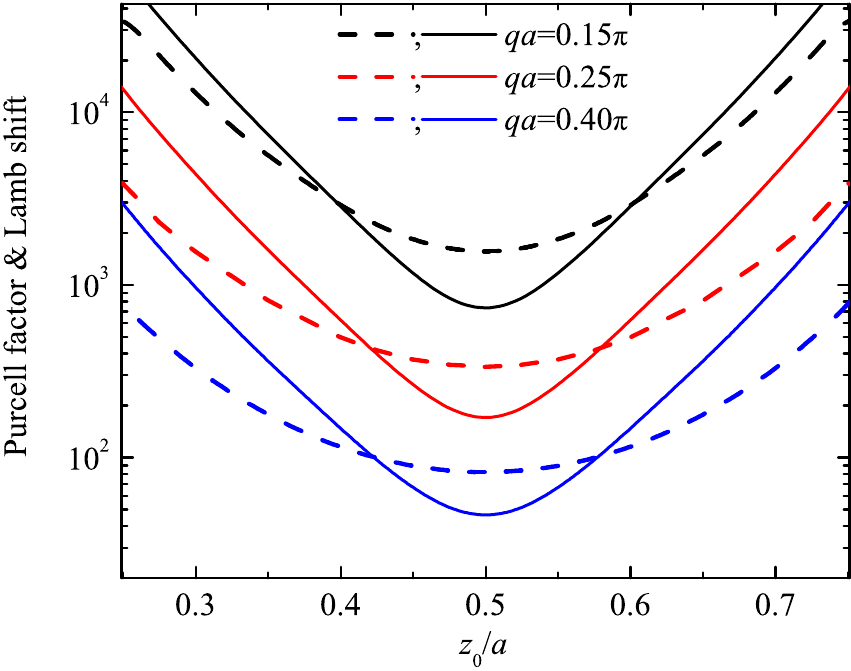}
\end{center}
 \caption{(Color online) Purcell factor (solid lines) and Lamb shift (dashed
lines) dependence on the coordinate $z_0$ of the source in the unit cell for
different values of $qa$. Calculation parameters  as the same as for
Fig.~\ref{fig:spat_hyp}.
}
\label{fig:lamb}
\end{figure}

The failure of the single dipole model in the hyperbolic medium is also revealed in the frequency
dependence of the Purcell factor, Fig.~\ref{fig:hyp_ell_q}a.  The dashed curve, calculated for single dipole, tends to the limit \eqref{eq:f_2b}, which is frequency independent.
However, the Purcell factor  in the hyperbolic medium diverges at low frequencies as $1/q^3$, according to
Eq.~\eqref{eq:f_div}. The Lamb shift $l$, calculated in hyperbolic medium for different values of $qa$ is presented in Fig.~\ref{fig:lamb} by the dashed curves.  Lamb shift is of the same order as the Purcell factor (dashed curve) and has similar near-field singularities at the node sites.

% % % % % % % % % % % % % % % % % % % % %

To summarize, Figs.~\ref{fig:hyp_ell}--\ref{fig:lamb} underline the importance of the
local-field effects in the hyperbolic medium and confirm the collective origin of the
spontaneous emission enhancement.

% % % % % % % % % % % % % % % % % % % % % % % % % % %

\section{Conclusions}\label{sec:concl}

We have developed the analytical theory of light-matter coupling in discrete
hyperbolic metamaterials in the framework of the discrete model of a cubic lattice 
of uniaxial resonant dipoles. We have calculated Purcell factor, Lamb shift, and Green
function for such a discrete model, and we have demonstrated that  the optimal emitter 
position is in the local field maxima, close to the lattice nodes. We have demonstrated
that the density of states is drastically enhanced in the hyperbolic regime as compared 
to other cases including vacuum, elliptic regime, or single resonant dipole case.
As a result, a huge number of lattice dipoles are efficiently excited by the
emitter, which has been visualized by calculating the Green function of
the lattice. The Green function has a shape of a conus: the field propagates along 
the directions close to symmetry  axis $z$ and decays in the $xy$  plane. 
Discrete character of the problem results in the strong spatial modulation of the Green function.

The calculated absolute values of the Purcell factor are rather challenging 
for the current realization of metamaterials. This is mainly due to the point dipole approximation we have utilized: as distance to the scatterers becomes comparable to their sizes, higher order multipoles must be accounted for. This  will inevitably reduce the local field and suppress the Purcell factor. Achieving large density of states enhancement described by Eq.~\eqref{eq:gain}  is also not easy. Finally, the losses are inevitable and can significantly influence the numerical answers. Nevertheless, we believe that our results will remain qualitatively correct for more complex settings, and they provide an important 
insight into the physics of hyperbolic metamaterials.

\acknowledgements{This work has been supported by the
Ministry of Education and Science of Russian Federation, the ``Dynasty'' Foundation,
 Russian Foundation for Basic Research, European project POLAPHEN, EPSRC~(UK),  and the Australian Research Council. The authors acknowledge useful discussions with C.R. Simovski.
}

%    \bibliography{all_cp1251b}

\begin{thebibliography}{47}
\expandafter\ifx\csname natexlab\endcsname\relax\def\natexlab#1{#1}\fi
\expandafter\ifx\csname bibnamefont\endcsname\relax
  \def\bibnamefont#1{#1}\fi
\expandafter\ifx\csname bibfnamefont\endcsname\relax
  \def\bibfnamefont#1{#1}\fi
\expandafter\ifx\csname citenamefont\endcsname\relax
  \def\citenamefont#1{#1}\fi
\expandafter\ifx\csname url\endcsname\relax
  \def\url#1{\texttt{#1}}\fi
\expandafter\ifx\csname urlprefix\endcsname\relax\def\urlprefix{URL }\fi
\providecommand{\bibinfo}[2]{#2}
\providecommand{\eprint}[2][]{\url{#2}}

\bibitem[{\citenamefont{{Jacob} and {Shalaev}}(2011)}]{shalaev2011}
\bibinfo{author}{\bibfnamefont{Z.}~\bibnamefont{{Jacob}}} \bibnamefont{and}
  \bibinfo{author}{\bibfnamefont{V.~M.} \bibnamefont{{Shalaev}}},
  \bibinfo{journal}{Science} \textbf{\bibinfo{volume}{334}},
  \bibinfo{pages}{463} (\bibinfo{year}{2011}).

\bibitem[{\citenamefont{Krishnamoorthy
  et~al.}(2012)\citenamefont{Krishnamoorthy, Jacob, Narimanov, Kretzschmar, and
  Menon}}]{Krishnamoorthy2012}
\bibinfo{author}{\bibfnamefont{H.~N.~S.} \bibnamefont{Krishnamoorthy}},
  \bibinfo{author}{\bibfnamefont{Z.}~\bibnamefont{Jacob}},
  \bibinfo{author}{\bibfnamefont{E.}~\bibnamefont{Narimanov}},
  \bibinfo{author}{\bibfnamefont{I.}~\bibnamefont{Kretzschmar}},
  \bibnamefont{and} \bibinfo{author}{\bibfnamefont{V.~M.} \bibnamefont{Menon}},
  \bibinfo{journal}{Science} \textbf{\bibinfo{volume}{336}},
  \bibinfo{pages}{205} (\bibinfo{year}{2012}).

\bibitem[{\citenamefont{{Gibbs} et~al.}(2011)\citenamefont{{Gibbs}, {Khitrova},
  and {Koch}}}]{khitrova2011}
\bibinfo{author}{\bibfnamefont{H.}~\bibnamefont{{Gibbs}}},
  \bibinfo{author}{\bibfnamefont{G.}~\bibnamefont{{Khitrova}}},
  \bibnamefont{and} \bibinfo{author}{\bibfnamefont{S.}~\bibnamefont{{Koch}}},
  \bibinfo{journal}{Nature Photonics} \textbf{\bibinfo{volume}{5}},
  \bibinfo{pages}{273} (\bibinfo{year}{2011}).

\bibitem[{\citenamefont{Vogel and Welsch}(2006)}]{Welsch2006}
\bibinfo{author}{\bibfnamefont{W.}~\bibnamefont{Vogel}} \bibnamefont{and}
  \bibinfo{author}{\bibfnamefont{D.-G.} \bibnamefont{Welsch}},
  \emph{\bibinfo{title}{Quantum Optics}} (\bibinfo{publisher}{Wiley},
  \bibinfo{address}{Weinheim}, \bibinfo{year}{2006}).

\bibitem[{\citenamefont{Ivchenko}(2005)}]{Ivchenko2005}
\bibinfo{author}{\bibfnamefont{E.~L.} \bibnamefont{Ivchenko}},
  \emph{\bibinfo{title}{Optical spectroscopy of semiconductor nanostructures}}
  (\bibinfo{publisher}{Alpha Science International}, \bibinfo{address}{Harrow,
  UK}, \bibinfo{year}{2005}).

\bibitem[{\citenamefont{Kavokin et~al.}(2006)\citenamefont{Kavokin, Baumberg,
  Malpuech, and Laussy}}]{kavbamalas}
\bibinfo{author}{\bibfnamefont{A.}~\bibnamefont{Kavokin}},
  \bibinfo{author}{\bibfnamefont{J.}~\bibnamefont{Baumberg}},
  \bibinfo{author}{\bibfnamefont{G.}~\bibnamefont{Malpuech}}, \bibnamefont{and}
  \bibinfo{author}{\bibfnamefont{F.}~\bibnamefont{Laussy}},
  \emph{\bibinfo{title}{Microcavities}} (\bibinfo{publisher}{Clarendon Press},
  \bibinfo{address}{Oxford}, \bibinfo{year}{2006}).

\bibitem[{\citenamefont{Silveirinha and Maslovski}(2012)}]{maslovski2012}
\bibinfo{author}{\bibfnamefont{M.~G.} \bibnamefont{Silveirinha}}
  \bibnamefont{and} \bibinfo{author}{\bibfnamefont{S.~I.}
  \bibnamefont{Maslovski}}, \bibinfo{journal}{Phys. Rev. B}
  \textbf{\bibinfo{volume}{85}}, \bibinfo{pages}{155125}
  (\bibinfo{year}{2012}).

\bibitem[{\citenamefont{Felsen and Marcuvitz}(2003)}]{Felsen}
\bibinfo{author}{\bibfnamefont{L.}~\bibnamefont{Felsen}} \bibnamefont{and}
  \bibinfo{author}{\bibfnamefont{N.}~\bibnamefont{Marcuvitz}},
  \emph{\bibinfo{title}{Radiation and scattering of waves}}
  (\bibinfo{publisher}{Wiley Interscience}, \bibinfo{address}{New York},
  \bibinfo{year}{2003}).

\bibitem[{\citenamefont{Lindell et~al.}(2001)\citenamefont{Lindell, Tretyakov,
  Nikoskinen, and Ilvonen}}]{lindell2001}
\bibinfo{author}{\bibfnamefont{I.~V.} \bibnamefont{Lindell}},
  \bibinfo{author}{\bibfnamefont{S.~A.} \bibnamefont{Tretyakov}},
  \bibinfo{author}{\bibfnamefont{K.~I.} \bibnamefont{Nikoskinen}},
  \bibnamefont{and} \bibinfo{author}{\bibfnamefont{S.}~\bibnamefont{Ilvonen}},
  \bibinfo{journal}{Microwave and Optical Technology Lett.}
  \textbf{\bibinfo{volume}{31}}, \bibinfo{pages}{129} (\bibinfo{year}{2001}).

\bibitem[{\citenamefont{Smith and Schurig}(2003)}]{Smith2003}
\bibinfo{author}{\bibfnamefont{D.~R.} \bibnamefont{Smith}} \bibnamefont{and}
  \bibinfo{author}{\bibfnamefont{D.}~\bibnamefont{Schurig}},
  \bibinfo{journal}{Phys. Rev. Lett.} \textbf{\bibinfo{volume}{90}},
  \bibinfo{pages}{077405} (\bibinfo{year}{2003}).

\bibitem[{\citenamefont{Fisher and Gould}(1969)}]{fisher1969}
\bibinfo{author}{\bibfnamefont{R.~K.} \bibnamefont{Fisher}} \bibnamefont{and}
  \bibinfo{author}{\bibfnamefont{R.~W.} \bibnamefont{Gould}},
  \bibinfo{journal}{Phys. Rev. Lett.} \textbf{\bibinfo{volume}{22}},
  \bibinfo{pages}{1093} (\bibinfo{year}{1969}).

\bibitem[{\citenamefont{Sun et~al.}(2011)\citenamefont{Sun, Zhou, Li, and
  Kang}}]{Sun2011}
\bibinfo{author}{\bibfnamefont{J.}~\bibnamefont{Sun}},
  \bibinfo{author}{\bibfnamefont{J.}~\bibnamefont{Zhou}},
  \bibinfo{author}{\bibfnamefont{B.}~\bibnamefont{Li}}, \bibnamefont{and}
  \bibinfo{author}{\bibfnamefont{F.}~\bibnamefont{Kang}},
  \bibinfo{journal}{Appl. Phys. Lett.} \textbf{\bibinfo{volume}{98}},
  \bibinfo{eid}{101901} (\bibinfo{year}{2011}).

\bibitem[{\citenamefont{Wurtz et~al.}(2008)\citenamefont{Wurtz, Dickson,
  O'Connor, Atkinson, Hendren, Evans, Pollard, and Zayats}}]{Wurtz2008}
\bibinfo{author}{\bibfnamefont{G.~A.} \bibnamefont{Wurtz}},
  \bibinfo{author}{\bibfnamefont{W.}~\bibnamefont{Dickson}},
  \bibinfo{author}{\bibfnamefont{D.}~\bibnamefont{O'Connor}},
  \bibinfo{author}{\bibfnamefont{R.}~\bibnamefont{Atkinson}},
  \bibinfo{author}{\bibfnamefont{W.}~\bibnamefont{Hendren}},
  \bibinfo{author}{\bibfnamefont{P.}~\bibnamefont{Evans}},
  \bibinfo{author}{\bibfnamefont{R.}~\bibnamefont{Pollard}}, \bibnamefont{and}
  \bibinfo{author}{\bibfnamefont{A.~V.} \bibnamefont{Zayats}},
  \bibinfo{journal}{Opt. Express} \textbf{\bibinfo{volume}{16}},
  \bibinfo{pages}{7460} (\bibinfo{year}{2008}).

\bibitem[{\citenamefont{Noginov et~al.}(2009)\citenamefont{Noginov, Barnakov,
  Zhu, Tumkur, Li, and Narimanov}}]{narimanov2009b}
\bibinfo{author}{\bibfnamefont{M.~A.} \bibnamefont{Noginov}},
  \bibinfo{author}{\bibfnamefont{Y.~A.} \bibnamefont{Barnakov}},
  \bibinfo{author}{\bibfnamefont{G.}~\bibnamefont{Zhu}},
  \bibinfo{author}{\bibfnamefont{T.}~\bibnamefont{Tumkur}},
  \bibinfo{author}{\bibfnamefont{H.}~\bibnamefont{Li}}, \bibnamefont{and}
  \bibinfo{author}{\bibfnamefont{E.~E.} \bibnamefont{Narimanov}},
  \bibinfo{journal}{Appl. Phys. Lett.} \textbf{\bibinfo{volume}{94}},
  \bibinfo{eid}{151105} (\bibinfo{year}{2009}).

\bibitem[{\citenamefont{Simovski et~al.}(2012)\citenamefont{Simovski, Belov,
  Atrashchenko, and Kivshar}}]{simovski2012}
\bibinfo{author}{\bibfnamefont{C.~R.} \bibnamefont{Simovski}},
  \bibinfo{author}{\bibfnamefont{P.~A.} \bibnamefont{Belov}},
  \bibinfo{author}{\bibfnamefont{A.~V.} \bibnamefont{Atrashchenko}},
  \bibnamefont{and} \bibinfo{author}{\bibfnamefont{Y.~S.}
  \bibnamefont{Kivshar}}, \bibinfo{journal}{Adv. Materials}
  (\bibinfo{year}{2012}), \bibinfo{note}{in press}.

\bibitem[{\citenamefont{Elser et~al.}(2007)\citenamefont{Elser, Podolskiy,
  Salakhutdinov, and Avrutsky}}]{elser2007b}
\bibinfo{author}{\bibfnamefont{J.}~\bibnamefont{Elser}},
  \bibinfo{author}{\bibfnamefont{V.~A.} \bibnamefont{Podolskiy}},
  \bibinfo{author}{\bibfnamefont{I.}~\bibnamefont{Salakhutdinov}},
  \bibnamefont{and} \bibinfo{author}{\bibfnamefont{I.}~\bibnamefont{Avrutsky}},
  \bibinfo{journal}{Appl. Phys. Lett.} \textbf{\bibinfo{volume}{90}},
  \bibinfo{eid}{191109} (pages~\bibinfo{numpages}{3}) (\bibinfo{year}{2007}).

\bibitem[{\citenamefont{Orlov et~al.}(2011)\citenamefont{Orlov, Voroshilov,
  Belov, and Kivshar}}]{orlov2011}
\bibinfo{author}{\bibfnamefont{A.~A.} \bibnamefont{Orlov}},
  \bibinfo{author}{\bibfnamefont{P.~M.} \bibnamefont{Voroshilov}},
  \bibinfo{author}{\bibfnamefont{P.~A.} \bibnamefont{Belov}}, \bibnamefont{and}
  \bibinfo{author}{\bibfnamefont{Y.~S.} \bibnamefont{Kivshar}},
  \bibinfo{journal}{Phys. Rev. B} \textbf{\bibinfo{volume}{84}},
  \bibinfo{pages}{045424} (\bibinfo{year}{2011}).

\bibitem[{\citenamefont{{Jacob} et~al.}(2010)\citenamefont{{Jacob}, {Kim},
  {Naik}, {Boltasseva}, {Narimanov}, and {Shalaev}}}]{narimanov2010}
\bibinfo{author}{\bibfnamefont{Z.}~\bibnamefont{{Jacob}}},
  \bibinfo{author}{\bibfnamefont{J.}~\bibnamefont{{Kim}}},
  \bibinfo{author}{\bibfnamefont{G.~V.} \bibnamefont{{Naik}}},
  \bibinfo{author}{\bibfnamefont{A.}~\bibnamefont{{Boltasseva}}},
  \bibinfo{author}{\bibfnamefont{E.~E.} \bibnamefont{{Narimanov}}},
  \bibnamefont{and} \bibinfo{author}{\bibfnamefont{V.~M.}
  \bibnamefont{{Shalaev}}}, \bibinfo{journal}{Appl. Phys. B: Lasers and Optics}
  \textbf{\bibinfo{volume}{100}}, \bibinfo{pages}{215} (\bibinfo{year}{2010}).

\bibitem[{\citenamefont{Noginov et~al.}(2010)\citenamefont{Noginov, Li,
  Barnakov, Dryden, Nataraj, Zhu, Bonner, Mayy, Jacob, and
  Narimanov}}]{Noginov2010}
\bibinfo{author}{\bibfnamefont{M.~A.} \bibnamefont{Noginov}},
  \bibinfo{author}{\bibfnamefont{H.}~\bibnamefont{Li}},
  \bibinfo{author}{\bibfnamefont{Y.~A.} \bibnamefont{Barnakov}},
  \bibinfo{author}{\bibfnamefont{D.}~\bibnamefont{Dryden}},
  \bibinfo{author}{\bibfnamefont{G.}~\bibnamefont{Nataraj}},
  \bibinfo{author}{\bibfnamefont{G.}~\bibnamefont{Zhu}},
  \bibinfo{author}{\bibfnamefont{C.~E.} \bibnamefont{Bonner}},
  \bibinfo{author}{\bibfnamefont{M.}~\bibnamefont{Mayy}},
  \bibinfo{author}{\bibfnamefont{Z.}~\bibnamefont{Jacob}}, \bibnamefont{and}
  \bibinfo{author}{\bibfnamefont{E.~E.} \bibnamefont{Narimanov}},
  \bibinfo{journal}{Opt. Lett.} \textbf{\bibinfo{volume}{35}},
  \bibinfo{pages}{1863} (\bibinfo{year}{2010}).

\bibitem[{\citenamefont{Ni et~al.}(2011)\citenamefont{Ni, Ishii, Thoreson,
  Shalaev, Han, Lee, and Kildishev}}]{Ni2011}
\bibinfo{author}{\bibfnamefont{X.}~\bibnamefont{Ni}},
  \bibinfo{author}{\bibfnamefont{S.}~\bibnamefont{Ishii}},
  \bibinfo{author}{\bibfnamefont{M.~D.} \bibnamefont{Thoreson}},
  \bibinfo{author}{\bibfnamefont{V.~M.} \bibnamefont{Shalaev}},
  \bibinfo{author}{\bibfnamefont{S.}~\bibnamefont{Han}},
  \bibinfo{author}{\bibfnamefont{S.}~\bibnamefont{Lee}}, \bibnamefont{and}
  \bibinfo{author}{\bibfnamefont{A.~V.} \bibnamefont{Kildishev}},
  \bibinfo{journal}{Opt. Express} \textbf{\bibinfo{volume}{19}},
  \bibinfo{pages}{25242} (\bibinfo{year}{2011}).

\bibitem[{\citenamefont{Kim et~al.}(2012)\citenamefont{Kim, Drachev, Jacob,
  Naik, Boltasseva, Narimanov, and Shalaev}}]{Kim2012}
\bibinfo{author}{\bibfnamefont{J.}~\bibnamefont{Kim}},
  \bibinfo{author}{\bibfnamefont{V.~P.} \bibnamefont{Drachev}},
  \bibinfo{author}{\bibfnamefont{Z.}~\bibnamefont{Jacob}},
  \bibinfo{author}{\bibfnamefont{G.~V.} \bibnamefont{Naik}},
  \bibinfo{author}{\bibfnamefont{A.}~\bibnamefont{Boltasseva}},
  \bibinfo{author}{\bibfnamefont{E.~E.} \bibnamefont{Narimanov}},
  \bibnamefont{and} \bibinfo{author}{\bibfnamefont{V.~M.}
  \bibnamefont{Shalaev}}, \bibinfo{journal}{Opt. Express}
  \textbf{\bibinfo{volume}{20}}, \bibinfo{pages}{8100} (\bibinfo{year}{2012}).

\bibitem[{\citenamefont{{Cortes} et~al.}(2012)\citenamefont{{Cortes}, {Newman},
  {Molesky}, and {Jacob}}}]{Cortes2012Arxiv}
\bibinfo{author}{\bibfnamefont{C.~L.} \bibnamefont{{Cortes}}},
  \bibinfo{author}{\bibfnamefont{W.}~\bibnamefont{{Newman}}},
  \bibinfo{author}{\bibfnamefont{S.}~\bibnamefont{{Molesky}}},
  \bibnamefont{and} \bibinfo{author}{\bibfnamefont{Z.}~\bibnamefont{{Jacob}}},
  \bibinfo{journal}{ArXiv e-prints}  (\bibinfo{year}{2012}),
  \eprint{1204.5529}.

\bibitem[{\citenamefont{{Jacob} et~al.}(2009)\citenamefont{{Jacob},
  {Smolyaninov}, and {Narimanov}}}]{narimanov2009}
\bibinfo{author}{\bibfnamefont{Z.}~\bibnamefont{{Jacob}}},
  \bibinfo{author}{\bibfnamefont{I.}~\bibnamefont{{Smolyaninov}}},
  \bibnamefont{and}
  \bibinfo{author}{\bibfnamefont{E.}~\bibnamefont{{Narimanov}}},
  \bibinfo{journal}{ArXiv e-prints}  (\bibinfo{year}{2009}),
  \eprint{0910.3981}.

\bibitem[{\citenamefont{Maslovski and Silveirinha}(2011)}]{maslovski2011}
\bibinfo{author}{\bibfnamefont{S.~I.} \bibnamefont{Maslovski}}
  \bibnamefont{and} \bibinfo{author}{\bibfnamefont{M.~G.}
  \bibnamefont{Silveirinha}}, \bibinfo{journal}{Phys. Rev. A}
  \textbf{\bibinfo{volume}{83}}, \bibinfo{pages}{022508}
  (\bibinfo{year}{2011}).

\bibitem[{\citenamefont{Iorsh et~al.}(2012)\citenamefont{Iorsh, Poddubny,
  Orlov, Belov, and Kivshar}}]{Iorsh2012}
\bibinfo{author}{\bibfnamefont{I.}~\bibnamefont{Iorsh}},
  \bibinfo{author}{\bibfnamefont{A.}~\bibnamefont{Poddubny}},
  \bibinfo{author}{\bibfnamefont{A.}~\bibnamefont{Orlov}},
  \bibinfo{author}{\bibfnamefont{P.}~\bibnamefont{Belov}}, \bibnamefont{and}
  \bibinfo{author}{\bibfnamefont{Y.~S.} \bibnamefont{Kivshar}},
  \bibinfo{journal}{Phys. Lett. A} \textbf{\bibinfo{volume}{376}},
  \bibinfo{pages}{185 } (\bibinfo{year}{2012}).

\bibitem[{\citenamefont{Xie et~al.}(2009)\citenamefont{Xie, Leung, and
  Tsai}}]{Xie2009}
\bibinfo{author}{\bibfnamefont{H.}~\bibnamefont{Xie}},
  \bibinfo{author}{\bibfnamefont{P.}~\bibnamefont{Leung}}, \bibnamefont{and}
  \bibinfo{author}{\bibfnamefont{D.}~\bibnamefont{Tsai}},
  \bibinfo{journal}{Solid State Comm.} \textbf{\bibinfo{volume}{149}},
  \bibinfo{pages}{625 } (\bibinfo{year}{2009}).

\bibitem[{\citenamefont{Kidwai et~al.}(2011)\citenamefont{Kidwai, Zhukovsky,
  and Sipe}}]{sipe2011}
\bibinfo{author}{\bibfnamefont{O.}~\bibnamefont{Kidwai}},
  \bibinfo{author}{\bibfnamefont{S.~V.} \bibnamefont{Zhukovsky}},
  \bibnamefont{and} \bibinfo{author}{\bibfnamefont{J.~E.} \bibnamefont{Sipe}},
  \bibinfo{journal}{Opt. Lett.} \textbf{\bibinfo{volume}{36}},
  \bibinfo{pages}{2530} (\bibinfo{year}{2011}).

\bibitem[{\citenamefont{{Yan} et~al.}(2012)\citenamefont{{Yan}, {Wubs}, and
  {Asger Mortensen}}}]{Yan2012arXiv}
\bibinfo{author}{\bibfnamefont{W.}~\bibnamefont{{Yan}}},
  \bibinfo{author}{\bibfnamefont{M.}~\bibnamefont{{Wubs}}}, \bibnamefont{and}
  \bibinfo{author}{\bibfnamefont{N.}~\bibnamefont{{Asger Mortensen}}},
  \bibinfo{journal}{ArXiv e-prints}  (\bibinfo{year}{2012}),
  \eprint{1204.5413}.

\bibitem[{\citenamefont{Poddubny et~al.}(2011)\citenamefont{Poddubny, Belov,
  and Kivshar}}]{poddubny2011pra}
\bibinfo{author}{\bibfnamefont{A.~N.} \bibnamefont{Poddubny}},
  \bibinfo{author}{\bibfnamefont{P.~A.} \bibnamefont{Belov}}, \bibnamefont{and}
  \bibinfo{author}{\bibfnamefont{Y.~S.} \bibnamefont{Kivshar}},
  \bibinfo{journal}{Phys. Rev. A} \textbf{\bibinfo{volume}{84}},
  \bibinfo{pages}{023807} (\bibinfo{year}{2011}).

\bibitem[{\citenamefont{{Ivchenko} et~al.}(2000)\citenamefont{{Ivchenko}, {Fu},
  and {Willander}}}]{Ivchenko2000}
\bibinfo{author}{\bibfnamefont{E.~L.} \bibnamefont{{Ivchenko}}},
  \bibinfo{author}{\bibfnamefont{Y.}~\bibnamefont{{Fu}}}, \bibnamefont{and}
  \bibinfo{author}{\bibfnamefont{M.}~\bibnamefont{{Willander}}},
  \bibinfo{journal}{Phys. Solid State} \textbf{\bibinfo{volume}{42}},
  \bibinfo{pages}{1756} (\bibinfo{year}{2000}).

\bibitem[{\citenamefont{Deutsch et~al.}(1995)\citenamefont{Deutsch, Spreeuw,
  Rolston, and Phillips}}]{deutsch1995}
\bibinfo{author}{\bibfnamefont{I.~H.} \bibnamefont{Deutsch}},
  \bibinfo{author}{\bibfnamefont{R.~J.~C.} \bibnamefont{Spreeuw}},
  \bibinfo{author}{\bibfnamefont{S.~L.} \bibnamefont{Rolston}},
  \bibnamefont{and} \bibinfo{author}{\bibfnamefont{W.~D.}
  \bibnamefont{Phillips}}, \bibinfo{journal}{Phys. Rev. A}
  \textbf{\bibinfo{volume}{52}}, \bibinfo{pages}{1394} (\bibinfo{year}{1995}).

\bibitem[{\citenamefont{{van Coevorden} et~al.}(1996)\citenamefont{{van
  Coevorden}, {Sprik}, {Tip}, and {Lagendijk}}}]{lagendijk1996}
\bibinfo{author}{\bibfnamefont{D.~V.} \bibnamefont{{van Coevorden}}},
  \bibinfo{author}{\bibfnamefont{R.}~\bibnamefont{{Sprik}}},
  \bibinfo{author}{\bibfnamefont{A.}~\bibnamefont{{Tip}}}, \bibnamefont{and}
  \bibinfo{author}{\bibfnamefont{A.}~\bibnamefont{{Lagendijk}}},
  \bibinfo{journal}{Phys. Rev. Lett.} \textbf{\bibinfo{volume}{77}},
  \bibinfo{pages}{2412} (\bibinfo{year}{1996}).

\bibitem[{\citenamefont{Kagan}(1999)}]{kagan1999}
\bibinfo{author}{\bibfnamefont{Y.}~\bibnamefont{Kagan}},
  \bibinfo{journal}{Hyperfine Interactions} \textbf{\bibinfo{volume}{123}},
  \bibinfo{pages}{83} (\bibinfo{year}{1999}).

\bibitem[{\citenamefont{{Purcell} and {Pennypacker}}(1973)}]{Purcell1973}
\bibinfo{author}{\bibfnamefont{E.~M.} \bibnamefont{{Purcell}}}
  \bibnamefont{and} \bibinfo{author}{\bibfnamefont{C.~R.}
  \bibnamefont{{Pennypacker}}}, \bibinfo{journal}{Astroph. J.}
  \textbf{\bibinfo{volume}{186}}, \bibinfo{pages}{705} (\bibinfo{year}{1973}).

\bibitem[{\citenamefont{Belov and Simovski}(2005)}]{belov2005}
\bibinfo{author}{\bibfnamefont{P.~A.} \bibnamefont{Belov}} \bibnamefont{and}
  \bibinfo{author}{\bibfnamefont{C.~R.} \bibnamefont{Simovski}},
  \bibinfo{journal}{Phys. Rev. E} \textbf{\bibinfo{volume}{72}},
  \bibinfo{pages}{026615} (\bibinfo{year}{2005}).

\bibitem[{\citenamefont{Silveirinha and Belov}(2008)}]{belov2008b}
\bibinfo{author}{\bibfnamefont{M.~G.} \bibnamefont{Silveirinha}}
  \bibnamefont{and} \bibinfo{author}{\bibfnamefont{P.~A.} \bibnamefont{Belov}},
  \bibinfo{journal}{Phys. Rev. B} \textbf{\bibinfo{volume}{77}},
  \bibinfo{pages}{233104} (\bibinfo{year}{2008}).

\bibitem[{\citenamefont{Korringa}(1947)}]{Korringa1947}
\bibinfo{author}{\bibfnamefont{J.}~\bibnamefont{Korringa}},
  \bibinfo{journal}{Physica} \textbf{\bibinfo{volume}{13}}, \bibinfo{pages}{392
  } (\bibinfo{year}{1947}).

\bibitem[{\citenamefont{{de Vries} et~al.}(1998)\citenamefont{{de Vries}, {van
  Coevorden}, and Lagendijk}}]{lagendijk_review}
\bibinfo{author}{\bibfnamefont{P.}~\bibnamefont{{de Vries}}},
  \bibinfo{author}{\bibfnamefont{D.~V.} \bibnamefont{{van Coevorden}}},
  \bibnamefont{and}
  \bibinfo{author}{\bibfnamefont{A.}~\bibnamefont{Lagendijk}},
  \bibinfo{journal}{Rev. Mod. Phys.} \textbf{\bibinfo{volume}{70}},
  \bibinfo{pages}{447} (\bibinfo{year}{1998}).

\bibitem[{\citenamefont{Landau and Lifshitz}(1974)}]{landau08}
\bibinfo{author}{\bibfnamefont{L.}~\bibnamefont{Landau}} \bibnamefont{and}
  \bibinfo{author}{\bibfnamefont{E.}~\bibnamefont{Lifshitz}},
  \emph{\bibinfo{title}{Electrodynamics of Continuous Media}}
  (\bibinfo{publisher}{Pergamon}, \bibinfo{address}{New York},
  \bibinfo{year}{1974}).

\bibitem[{\citenamefont{Pollard et~al.}(2009)\citenamefont{Pollard, Murphy,
  Hendren, Evans, Atkinson, Wurtz, Zayats, and Podolskiy}}]{pollard2009}
\bibinfo{author}{\bibfnamefont{R.~J.} \bibnamefont{Pollard}},
  \bibinfo{author}{\bibfnamefont{A.}~\bibnamefont{Murphy}},
  \bibinfo{author}{\bibfnamefont{W.~R.} \bibnamefont{Hendren}},
  \bibinfo{author}{\bibfnamefont{P.~R.} \bibnamefont{Evans}},
  \bibinfo{author}{\bibfnamefont{R.}~\bibnamefont{Atkinson}},
  \bibinfo{author}{\bibfnamefont{G.~A.} \bibnamefont{Wurtz}},
  \bibinfo{author}{\bibfnamefont{A.~V.} \bibnamefont{Zayats}},
  \bibnamefont{and} \bibinfo{author}{\bibfnamefont{V.~A.}
  \bibnamefont{Podolskiy}}, \bibinfo{journal}{Phys. Rev. Lett.}
  \textbf{\bibinfo{volume}{102}}, \bibinfo{pages}{127405}
  (\bibinfo{year}{2009}).

\bibitem[{\citenamefont{Belov et~al.}(2003)\citenamefont{Belov, Marqu\'es,
  Maslovski, Nefedov, Silveirinha, Simovski, and Tretyakov}}]{belov2003}
\bibinfo{author}{\bibfnamefont{P.~A.} \bibnamefont{Belov}},
  \bibinfo{author}{\bibfnamefont{R.}~\bibnamefont{Marqu\'es}},
  \bibinfo{author}{\bibfnamefont{S.~I.} \bibnamefont{Maslovski}},
  \bibinfo{author}{\bibfnamefont{I.~S.} \bibnamefont{Nefedov}},
  \bibinfo{author}{\bibfnamefont{M.}~\bibnamefont{Silveirinha}},
  \bibinfo{author}{\bibfnamefont{C.~R.} \bibnamefont{Simovski}},
  \bibnamefont{and} \bibinfo{author}{\bibfnamefont{S.~A.}
  \bibnamefont{Tretyakov}}, \bibinfo{journal}{Phys. Rev. B}
  \textbf{\bibinfo{volume}{67}}, \bibinfo{pages}{113103}
  (\bibinfo{year}{2003}).

\bibitem[{\citenamefont{Savchenko and Savchenko}(2005)}]{savchenko2005}
\bibinfo{author}{\bibfnamefont{A.}~\bibnamefont{Savchenko}} \bibnamefont{and}
  \bibinfo{author}{\bibfnamefont{O.}~\bibnamefont{Savchenko}},
  \bibinfo{journal}{Technical Phys.} \textbf{\bibinfo{volume}{50}},
  \bibinfo{pages}{1366} (\bibinfo{year}{2005}).

\bibitem[{\citenamefont{C.Kittel}(1996)}]{KittelIntro}
\bibinfo{author}{\bibnamefont{C.Kittel}}, \emph{\bibinfo{title}{Introduction to
  Solid State Phys.}} (\bibinfo{publisher}{Wiley}, \bibinfo{address}{New York},
  \bibinfo{year}{1996}).

\bibitem[{\citenamefont{Swendsen and Callen}(1972)}]{Swendsen1972}
\bibinfo{author}{\bibfnamefont{R.~H.} \bibnamefont{Swendsen}} \bibnamefont{and}
  \bibinfo{author}{\bibfnamefont{H.}~\bibnamefont{Callen}},
  \bibinfo{journal}{Phys. Rev. B} \textbf{\bibinfo{volume}{6}},
  \bibinfo{pages}{2860} (\bibinfo{year}{1972}).

\bibitem[{\citenamefont{Toma\ifmmode~\check{s}\else \v{s}\fi{} and
  Lenac}(1999)}]{tomas1999}
\bibinfo{author}{\bibfnamefont{M.~S.} \bibnamefont{Toma\ifmmode~\check{s}\else
  \v{s}\fi{}}} \bibnamefont{and}
  \bibinfo{author}{\bibfnamefont{Z.}~\bibnamefont{Lenac}},
  \bibinfo{journal}{Phys. Rev. A} \textbf{\bibinfo{volume}{60}},
  \bibinfo{pages}{2431} (\bibinfo{year}{1999}).

\bibitem[{\citenamefont{Wang et~al.}(2004)\citenamefont{Wang, Kivshar, and
  Gu}}]{kivshar2004}
\bibinfo{author}{\bibfnamefont{X.-H.} \bibnamefont{Wang}},
  \bibinfo{author}{\bibfnamefont{Y.~S.} \bibnamefont{Kivshar}},
  \bibnamefont{and} \bibinfo{author}{\bibfnamefont{B.-Y.} \bibnamefont{Gu}},
  \bibinfo{journal}{Phys. Rev. Lett.} \textbf{\bibinfo{volume}{93}},
  \bibinfo{pages}{073901} (\bibinfo{year}{2004}).

\bibitem[{\citenamefont{Novotny and Hecht}(2006)}]{Novotny2006}
\bibinfo{author}{\bibfnamefont{L.}~\bibnamefont{Novotny}} \bibnamefont{and}
  \bibinfo{author}{\bibfnamefont{B.}~\bibnamefont{Hecht}},
  \emph{\bibinfo{title}{Principles of Nano-Optics}}
  (\bibinfo{publisher}{Cambridge University Press}, \bibinfo{address}{New
  York}, \bibinfo{year}{2006}).

\end{thebibliography}

\end{document}